\documentclass[11pt,preprint,showpacs,preprintnumbers,amssymb]{revtex4}
\usepackage{epsfig}
\usepackage{graphicx}
\usepackage{here}
\usepackage{dcolumn}
\usepackage{bm}

\evensidemargin=\oddsidemargin

\newcommand{\beq}{\begin{equation}}
\newcommand{\eeq}{\end{equation}}
\newcommand{\bq}{\begin{equation}}
\newcommand{\eq}{\end{equation}}
\newcommand{\ba}{\begin{array}}
\newcommand{\ea}{\end{array}}
\newcommand{\beqa}{\begin{eqnarray}}
\newcommand{\eeqa}{\end{eqnarray}}

\newcommand{\mev}{{\rm MeV}}
\newcommand{\gev}{${\rm GeV}$}

\def\O{{\cal O}}

\def\End{\end{document}}

\def\to{\rightarrow}
\def\To{\Rightarrow}

\def\dis{\displaystyle}
\def\f{\frac}
\def\ov{\overline}
\def\[{\left[}
\def\]{\right]}
\def\({\left(}
\def\){\right)}

\def\ra{\rightarrow}

\def\G{{\cal G}}
\def\GSM{{\cal G}_{\rm SM}}

\def\l{{\ell}}

\def\W{\mathcal W}
\def\Z{\mathcal Z}
\def\DL{\Delta {\mathcal L}}
\def\a{\alpha}
\def\b{\beta}
\def\GA{\Gamma}
\def\C{\mathcal C}
\def\U1EM{U(1)_{\rm em}}
\def\O{\mathcal O}
\def\R{\mathcal R}
\def\la{\langle}
\def\ra{\rangle}
\def\deg{\circ}
\def\leqq{\leqslant}

\def\cw{c_{\rm w}}
\def\sw{s_{\rm w}}

\def\[{\left[}
\def\]{\right]}
\def\dis{\displaystyle}

\def\cut{\Lambda}
\def\LL{\mathcal L}

\def\thisday{June, 2002}
\begin{document}

\setcounter{footnote}{1}
\renewcommand{\thefootnote}{\fnsymbol{footnote}}

 

\preprint{{\large hep-ph/0206056}}
\preprint{{\small JLAB-THY-02-20}} 
\preprint{{\small MADPH-02-1263}} 
\preprint{{\small UT-HEP-02-03}} 
\preprint{{\small WM-02-105} \hspace*{1.5cm}}


\title{
\vspace*{1cm}
{\Large ${\mbox {\boldmath  ${\tau-\mu}$}}$~
Flavor Violation as a Probe of the Scale of New Physics }
}%

\author{\sc {\large Deirdre Black}\,\footnote{
Electronic address: dblack@jlab.org}
}
\affiliation{%
Theory Group, Thomas Jefferson National Accelerator Facility,
Newport News, Virginia 23606 
}%

\author{\sc {\large Tao Han}\,\footnote{
Electronic address: than@pheno.physics.wisc.edu
}}
\affiliation{%
Department of Physics, 
University of Wisconsin, Madison, Wisconsin 53706
}%

\author{\sc {\large Hong-Jian He}\,\footnote{
Electronic address: hjhe@physics.utexas.edu}
}
\affiliation{%
Center for Particle Physics,
University of Texas at Austin, Texas 78712
}%

\author{\sc {\large Marc Sher}\,\footnote{
Electronic address: sher@physics.wm.edu}
}
\affiliation{%
Nuclear and Particle Theory Group,~College of William and Mary, 
Williamsburg, Virginia 23187
}%

\date{\thisday}

\begin{abstract}
\vspace*{2mm}
\noindent
Motivated by the recent strong experimental evidence of large 
$\nu_\mu-\nu_\tau$ neutrino mixing, we explore
current bounds on the analogous mixing in the charged lepton sector.  
We present a general formalism for dimension-6  fermionic effective 
operators involving
$\tau-\mu$ mixing with typical Lorentz structure
$(\ov{\mu} \,\Gamma \tau)(\ov{q}^\a \Gamma {q}^\b )$, and discuss 
their relationship to the standard model gauge symmetry
and the underlying flavor dynamics.  
We derive the low-energy constraints 
on the new physics scale associated with each operator, 
mostly from current experimental bounds 
on rare decay processes of $\tau$, hadrons or heavy quarks.
For operators involving at least one light quark ($u,d,s$), 
these constraints typically give a bound on the new physics scale of a few
TeV or higher. Those operators with two heavy quarks turn out 
to be more weakly constrained at
the present, giving bounds of  a few hundred GeV.
A few scalar and pseudo-scalar operators  
are free from all current experimental constraints. 
\end{abstract}


\maketitle

\baselineskip20pt   
\newpage
\renewcommand{\baselinestretch}{0.95}
\setcounter{footnote}{0}
\renewcommand{\thefootnote}{\arabic{footnote}}
\setcounter{page}{2}

\addtolength{\topmargin}{-4mm}
\setcounter{page}{2}  
\section{Introduction}

The mystery of ``flavor'' poses a major challenge in particle physics.
The unpredicted masses and mixings of three families of leptons and quarks 
in the flavor sector compose thirteen out of nineteen 
free parameters in the whole Standard Model (SM). 
Weak scale supersymmetry (SUSY) \cite{susy} as a leading candidate for 
new physics beyond the SM provides no further understanding 
about the origin of flavor. In fact, it extends the mystery of flavor
by necessarily adding three families of squarks and sleptons.
Without additional assumptions for flavor structure of the soft SUSY
breaking, supersymmetric theories often encounter phenomenolgical
difficulties, known as the SUSY flavor problem \cite{susyf}.
In dynamical models of  electroweak symmetry 
breaking, the phenomenological constraints
on flavor-changing-neutral-currents (FCNC) make 
it hard for model building to accommodate the observed heavy top quark mass, 
unless new dynamics associated with the top quark 
is introduced \cite{Hill-Rept}.
Predicting the full mass spectrum and mixing pattern in the flavor
sector may have to invoke new physics scales ranging from
the weak scale up to very high scales in a single unified theory.
Less ambitious and more practical approaches follow a ``bottom-up'' path, which effectively parametrize the new physics with 
flavor in a way not explicitly invoking unknown dynamics at the high scales.
For instance, certain ansatze for fermion masses matrices 
were advocated \cite{ansatz}, and realizations of 
horizontal symmetries were proposed\,\cite{FN,NS} to explain the fermion
mass hierarchies and Cabibbo-Kobayashi-Maskawa (CKM) mixings \cite{CKM}
at relatively low scales.
The FCNC fermionic Yukawa couplings 
to the Higgs sector can also be constructed 
in a phenomenologically viable way \cite{ansatzhiggs,hall}.
When the electroweak symmetry breaking sector is nonlinearly
realized \cite{CCWZ,weinberg,App}, new FCNC gauge interactions of fermions
can be economically described 
by effective operators of dimension-4 \cite{PZ,BU}; while for the
linearly realized Higgs sector, these couplings arise from dimension-6 
effective operators \cite{BW}.

The recent exciting evidence for neutrino oscillations \cite{atm,sol}
strongly points to nonzero neutrino masses with large (rather
than small) mixings which further discriminate the lepton sector from
the quark sector and deepens the mystery of ``flavor''. 
In fact, the atmospheric oscillation data \cite{atm}
favors maximal mixing between the $\mu$ and $\tau$ neutrinos\footnote{
The atmospheric mixing angle for $\nu_\mu\to \nu_\tau$ transition is measured
to be $31^\deg \leqq \theta_{\rm atm} \leqq 59^\deg$ at 
99\%\,C.L., with a central value   
$\theta_{\rm atm} \simeq 45^\deg$ \cite{atm}.}
via the Maki-Nakagawa-Sakata (MNS) matrix \cite{MNS}\footnote{
The MNS mixing matrix in the lepton charged current
is the analogue of CKM
in the quark charged current, but they exhibit completely different
structures.}.
This development has led to a great amount of theoretical effort,
with the hope of revealing the underlying new physics in the 
leptonic flavor sector \cite{numodel}.

In this paper, we systematically explore the
low energy constraints on effective 
operators induced by large $\tau-\mu$ mixing. This is
strongly motivated by the neutrino oscillation data,
in particular, the favored maximal mixing between the 
second and third generations via MNS matrix.
Note that the large or maximal mixings in the MNS matrix
may come from either neutrino mass
diagonalization, or lepton mass diagonalization, or from
both of them.
It is indeed tempting to search for charged lepton flavor violations,
in addition to the existing neutrino oscillation experiments. 
There are two classes of structures that can lead to flavor mixing
for charged leptons. The first class is that there is no
tree-level flavor mixing for charged leptons after mass
diagonalization. This is an analogue to the SM quark sector,
where the flavor mixing effects appear at tree level
only in the charged current sector via the CKM mixing matrix.
For instance, in a SM-like framework 
with additional mass and mixing parameters
in the neutrino sector, $\tau-\mu$ mixing is generated
at one-loop level and is generally suppressed by a factor of
$m_\nu^2/M_W^2$, which would be negligible. 
The second class yields tree-level mixing effects after
the mass diagonalization from the flavor eigenbasis into
the mass eigenbasis.
Typical theories in this class include extended models with 
extra $Z'$ or Higgs doublets, generic weak-scale SUSY models,
and dynamical models with compositeness,
which often have rich structures of flavor mixing, leading 
to testable new phenomena. By contemplating on the large lepton
mass hierarchy, dimensional analysis suggests the new physics 
scale associated with the mass and mixing of the third family 
leptons to be the lowest one in the lepton sector.

Before experiments can directly access the new physics
scale associated with the lepton flavor sector,
we use an effective theory formulation, obtained by
integrating out the heavy degrees of freedom from a  
more fundamental theory.
In Sec.\,II, we present our general formalism for the dimension-6
fermionic effective operators involving $\tau-\mu$ flavor mixing 
with typical Lorentz structures
\begin{equation}
\label{eq:4F-def}
\( \bar\mu\, \Gamma\, \tau \)
( \bar{q}^\a \,\Gamma\, {q}^\b ) \,,
\end{equation}
where $\Gamma$ contains possible Dirac $\gamma$-matrices.
This is beyond the simplest flavor-diagonal form of the
effective four-Fermi contact interactions \cite{Peskin}.
With the operators in Eq.\,(\ref{eq:4F-def}),
we analyze their relationship to the realizations of the SM 
gauge symmetry and the underlying flavor dynamics. We further
estimate the expected size of their coefficients, 
for a given cutoff scale at which the effective theory
breaks down and new physics sets in. 
We also comment on to what extent our formalism
can be applicable to loop-induced processes.
In Sec.\,III, we systematically explore the constraints 
on the new physics scale associated with each operator, mostly 
from current experimental bounds on rare decays of $\tau$, hadrons or heavy
quarks. For the operators involving at least one light quark ($u,d,s$), 
the current low energy constraints typically push the new physics scale 
to a few TeV or higher. Those operators with two heavy quarks are 
at the present subject to weaker constraints, only about a few hundred GeV.
Some of the scalar and pseudo-scalar operators are free from 
any experimental constraint so far. 
We summarize our results in Table\,I and conclude in Sec.\,IV.
Appendix\,A 
analyzes the relationship between our operators introduced
in Sec.\,II and those with explicit SM gauge symmetry in both
linear and non-linear realizations. 
In Appendix\,B, we further extend the above 
Eq.\,(\ref{eq:4F-def}) to include the forms with generic
lepton bilinear $\bar{\l}^\a \,\Gamma\, {\l}^\b $, and
derive the corresponding constraints from rare $\tau$ decays.

\section{Formalism for Effective Operators with 
         $\mbox{\boldmath $\tau \!-\! \mu$}$ Flavor Violation}

\subsection{\large Constructing the Dimension-6 
            \,$\mbox{\boldmath $\tau \!-\! \mu$}$\, Operators}

We consider an effective theory below the new physics scale
$\cut$, which can be generally defined as 
\beq
\label{eq:Leff}
\LL_{\rm eff} = \LL_{\rm SM} + \Delta\LL
\eeq
where $\LL_{\rm SM}$ is the SM Lagrangian density and $\Delta\LL$ denotes
the new physics contribution via effective operators. For the
current study, we will focus on the dimension-6 operators involving
third and second family leptons $(\tau,\,\mu)$,
\beq
\label{eq:taumu6}
\DL = 
\DL^{(6)}_{\tau\mu} 
=\dis\sum_{j,\alpha,\beta}
\f{~\C^j_{\alpha\beta}~}{\cut^2}
\left( \ov{\mu} ~\Gamma_j\, \tau \right) 
\left( \ov{q}^\alpha \,\Gamma_j\, q^\beta \right)
\,+\,{\rm H.c.}\,,
\eeq
where 
$\GA_j\in 
\(1,\,\gamma_5^{~},\,\gamma_\sigma^{~},\,\gamma_\sigma^{~}\gamma_5^{~}\)$ 
denotes relevant Dirac matrices, specifying scalar, pseudoscalar,
vector and axial vector couplings, respectively.
We will not consider the possibility that 
$\GA$ has tensor structure due to the following reasons:
(a) there are no two-body decays involving the tau with tensor
structure, and thus any bound would be extremely weak; 
(b) the tensor matrix elements involving quark bilinears are either
unknown or known very poorly, and thus bounds would be not
only very weak, but also rather uncertain; 
(c) the tensor structure does not generically appear in most 
models that we know.
Here, we consider  $\GA_j$  to be the same for both 
$\tau-\mu$ and $q^\a - q^\b$ bilinears, 
where $\alpha$ and $\beta$ run over all allowed combinations of quark flavors.
Under these considerations, we can show that Eq.\,(\ref{eq:taumu6})
is the most general form (containing one $\tau-\mu$ bilinear and 
one quark-bilinear) which respects the unbroken
$U(1)_{\rm em}$ gauge symmetry. 
As shown in the Appendix\,A, 
the operator Eq.\,(\ref{eq:taumu6}) corresponds to a nonlinear
realization of the electroweak gauge symmetry under which all fields
feel only the unbroken $U(1)_{\rm em}$.
For the linearly realized electroweak gauge symmetry where the
physical Higgs and would-be Goldstone bosons form the usual
Higgs doublet, the dimension-6 $\tau-\mu$ operators
have to respect the full electroweak gauge group
$\G_{\rm SM} = SU(2)_L\otimes U(1)_Y$  
and are thus found to have a more restricted
form,
\beq
\label{eq:taumu6-lin}
\DL^{\rm linear} = 
\DL^{{\rm linear}}_{\tau\mu (6)} 
=\dis\sum_{\alpha,\beta,\chi,\zeta}^j
\f{~\C^{j\chi\zeta}_{\alpha\beta}~}{\cut^2}
\left( \ov{\mu_\chi} ~\Gamma_j\, \tau_\chi \right) 
\left( \ov{q}_{\zeta}^{\alpha} \,\Gamma_j\, q_{\zeta}^{\beta} \right)
\,+\,{\rm H.c.}\,,
\eeq
where the chirality indices $\chi,\zeta = L,R$.
This restricts $\GA \in (\gamma^{~}_\sigma,\, \gamma_\sigma^{~}\gamma^{~}_5)$,
so that Eq.\,(\ref{eq:taumu6-lin}) only belongs to a sub-set of operators in
Eq.\,(\ref{eq:taumu6}).\footnote{Note that, in Eq.\,(\ref{eq:taumu6-lin}),
after combining the chirality
projection operators $P_{L,R}=(1\mp\gamma_5)/2$ 
with $\GA_j$'s, we can have structures
with different $\GA_j$ in each fermion-bilinear, which will not
be considered further [similar to the restriction we have added 
in Eq.\,(\ref{eq:taumu6})].}\,  
At first sight, this is somewhat surprising as the scalar and
pseudoscalar operators in Eq.\,(\ref{eq:taumu6}) are fully absent at dimension-6. However, it is interesting to note that such
scalar and pseudoscalar structures reappear in the dimension-8 
effective operators,
\beq
\label{eq:taumu8-lin}
\ba{ll}
\DL^{{\rm linear}}_{\tau\mu (8)} \!\!\! 
& = \dis\f{1}{\cut^4} \!\!\! 
\dis\sum_{\alpha,\beta,\a',\b'}^j   \!\!\! 
\C^{j(8)}_{\a\b\a'\b'}\!
\left( \ov{L}_\ell^{\a'}\Phi\Gamma_j\, \ell_R^{\b'} \right) \!\!
\left( \ov{L}_q^{\a}\Phi'\Gamma_j\, q_R^{\b} \right)\!
+\!
\widetilde{\C}^{j(8)}_{\a\b\a'\b'}\!
\left( \ov{L}_\ell^{\a'}\Phi\Gamma_j\, \ell_R^{\b'} \right) \!\! 
\left( \ov{q_R}^{\b}\Phi^{\prime\dag}\Gamma_j\, L_q^{\a} \right)\!
+\! {\rm H.c.}
\\[6mm]
&  \To \dis \sum_{\alpha,\beta}^j
\f{v^2\widehat{\C}^{j(8)}_{\a\b}}{2\cut^4}
\left( \ov{\mu} ~\Gamma_j\, \tau \right) 
\left( \ov{q}^{\alpha} \,\Gamma_{j}\, q^{\beta} \right)
\,+\,{\rm H.c.}\,, ~~~~~\(\,{\rm for}~\Phi=\langle\Phi\rangle\,\),
\ea
\eeq
where $\GA_j \in (1,\,\gamma_5)$, $L_\ell\,(L_q)$ is the 
left-handed lepton (quark) doublet, 
and $\Phi$ is the Higgs doublet
with hypercharge $\f{1}{2}$ and vacuum expectation value (VEV)
$\langle\Phi\rangle = v/\sqrt{2}\(0,\,1\)^T$. 
We also denote, in Eq.\,(\ref{eq:taumu8-lin}),
$\Phi' = \Phi$ for $q_R=d_R^{~}$ and 
$\Phi' = i\tau_2\Phi^\ast $ for $q_R=u_R^{~}$. 
Note that when the Higgs doublet takes its VEV, the dimension-8
operators in Eq.\,(\ref{eq:taumu8-lin})
reduce to the generic dimension-6 form
in Eq.\,(\ref{eq:taumu6-lin}) but with coefficients further
suppressed by an extra factor of $v^2/(2\cut^2)$.
Thus, they can be neglected in comparison with the leading
dimension-6 operators Eq.\,(\ref{eq:taumu6-lin}).
This means that for the linear realization of the SM gauge symmetry,
only vector and axial-vector operators with 
$\GA_j\in (\gamma_\sigma,\,\gamma_\sigma\gamma_5)$ 
are relevant at dimension-6 level.
For the phenomenological analysis in Sec.\,III, 
we will focus on analyzing
the bounds for the most general form in Eq.\,(\ref{eq:taumu6})
since Eq.\,(\ref{eq:taumu6})
contains the restricted form in Eq. (\ref{eq:taumu6-lin})
as a special case.

Finally, we note that in principle,
we could also include purely leptonic dimension-6 operators 
of the form 
$\( \bar\mu\, \Gamma\, \tau \)( \ov{\l}^a\, \Gamma\, \l^\b )$,
which contains an additional lepton bilinear 
$(\ov{\l}^a\, \Gamma\, \l^\b )$
instead of quark bilinear  $(\ov{q}^a\, \Gamma\, q^\b )$.
They may involve similar flavor 
dynamics as the effective operators
(\ref{eq:taumu6}), but unlike (\ref{eq:taumu6})
they are relevant to only a few low energy constraints. 
Most nontrivial bounds come from certain three-body rare $\tau$ decays
and all appear similar. 
We will summarize these separately in Appendix\,B.

\subsection{\large Theoretical Consideration for the Size of Coefficients}

The precise value of the dimensionless coefficient 
$\C^j_{\a\b}$ in Eq.\,(\ref{eq:taumu6}) should be derived from the
corresponding underlying theory in principle. 
In the current effective theory analysis, 
we will invoke a power counting estimate. As shown in the Appendix\,A,
the operator Eq.\,(\ref{eq:taumu6}) is formulated under 
the nonlinear realization of the SM gauge group $\GSM$, 
which provides a natural effective description 
of the strongly coupled electro-weak symmetry breaking
(EWSB) sector and/or compositeness. 
In this scenario, the natural size of $\C^j_{\a\b}$ for an effective
dimension-6 four-Fermi operator such as (\ref{eq:taumu6})
can be typically estimated as\,\cite{Hill-Rept},
\beq
\label{eq:C-def} 
\C^j_{\a\b}
            ~=~ 4\pi \,\O(1)   \,,  ~~~~~({\rm default}),
\eeq
which corresponds to an underlying theory with
a strong gauge coupling $\a_S^{~} = g_S^2/4\pi = \O(1)$.
Naive dimensional analysis (NDA) \cite{georgi1,georgi2}
provides another way to estimate operators in the nonlinear
realization. For the dimension-6 operators in Eq.\,(\ref{eq:taumu6}),
the NDA gives
\beq
\label{eq:C-NDA} 
\C_{\a\b}^j[{\rm NDA}]  
                       ~\lesssim~ (4\pi)^2 \,\O(1)   \,.
\eeq
which corresponds to an underlying theory with
a strong gauge coupling\footnote{
In our current study, we will assume that $\a_S^{~}$ is always below its
critical value so that there is no condensate formation for
$(\ov{\tau}\mu)$ and $(\ov{q}_\a^{~} q_\b^{~})$ channels.}\,
~$\a_S = g_S^2/4\pi \lesssim \O(4\pi)$\,.
So, in general, for the nonlinearly realized effective theory, we expect
$\C^j_{\a\b} \gtrsim 4\pi$,\, and to be conservative we will choose
the estimate Eq. (\ref{eq:C-def}) as the ``default'' value of our analysis.
With Eq.\,(\ref{eq:C-def}),
all the phenomenological bounds derived in the next section can be translated
into  bounds on the new physics scale $\cut$. The bounds with a
different counting of $\C^j_{\a\b} $, such as Eq.\,(\ref{eq:C-NDA})
above and   Eq.\,(\ref{eq:C-Weak}) below,
can be directly obtained from our
default results by simple rescaling.

In the weakly coupled theories, we have 
\beq
\label{eq:C-Weak} 
\C_{\a\b}^j  ~=~  \O(1) \,, ~~~~~({\rm weakly~coupled~scenario}).
\eeq
Since the new physics scale of a weakly coupled scenario is likely to be 
first determined by discovering the light new particles
(such as light Higgs boson(s) and 
a few lower-lying states of superpartners), 
we may mainly motivate the current analysis  
by the strongly coupled theories where
the estimate in  Eq.\,(\ref{eq:C-def}) or (\ref{eq:C-NDA}) 
can sensibly apply. 
However, as will be clear from the Sec.\,III and IV below,
even for the weakly coupled theories with 
Eq.\,(\ref{eq:C-Weak}), significant bounds on the new physics scale
$\cut$ can still be derived for the operators with the quark-bilinear
$\bar{q}^\a\,\GA\,q^\b$ containing no $t$ quarks, no $c$ quarks and at
most one $b$ quark.

The linear realization is more appropriate when there is a Higgs boson 
with a mass well below the scale $4\pi v\simeq 3.1$ TeV,
such as in the typical models with supersymmetry \cite{susy}
or composite Higgs models with Top-color \cite{Hill-Rept}\footnote{For
the minimal Top-seesaw models\,\cite{Hill-Rept}, 
the composite Higgs mass is generally
in the range around  $0.4\!-\!1$\,TeV\,\cite{seesaw,seesaw1}.}.
Finally, we note that the above estimates for the coefficient
are flavor-blind, i.e., we do not worry about the possible suppression
from flavor-violation effects. Below we will analyze 
how the large leptonic $\tau-\mu$ flavor mixing is naturally realized
in typical scenarios without further suppression.

\begin{figure}[H]
\begin{center} 
\includegraphics[width=7cm,height=4cm]{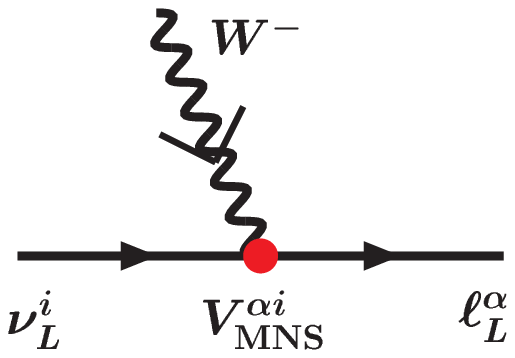} 
\includegraphics[width=7cm,height=4cm]{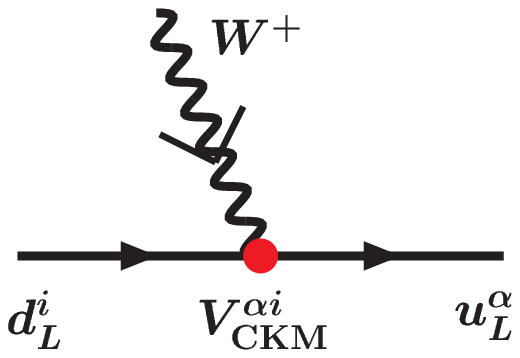}
\caption{The charged current interactions 
for lepton sector involving MNS mixings (left plot), and
for quark sector involving CKM mixings (right plot). }
\end{center}
\end{figure}

\subsection{{\large Neutrino Oscillations, 
            Large Lepton Mixings
            and \,$\mbox{\boldmath $\tau\!-\!\mu$}$\, 
            Flavor Violation Operators}}

Neutrino oscillation experiments can measure the
lepton charged current interactions involving (large) MNS mixings
(cf. left plot in Fig.\,1)
analogous to the quark charged current interactions
involving (small) CKM mixings (cf. right plot in Fig.\,1).
Thus we can generally write,
\beq
{\mathcal L}^{\ell}_{CC} =\dis
-\f{g}{\sqrt{2}}
\sum^{\a=e,\mu,\tau}_{i=1,\cdots,N}
\ov{\ell_L}^{\a}\gamma^\rho \,V_{\a i}\,\nu_{L}^{i}W^-_\rho \,+\, {\rm H.c.}
\eeq
where $V$ is the $3\times N$ unitary MNS mixing matrix containing
a product of two left-handed rotation matrices, 
\beq
\label{eq:MNS}
V_{\a i} = \dis
\sum_{\beta =1}^3 U^{\ell\,\ast}_{L\,\b\a} U^\nu_{L\,\b i}  ~,
\eeq 
where $U^{\ell}_L$ ($3\times3$) 
is from the lepton mass diagonalization
~$U^{\ell\,\dag}_L M_\ell U^{\ell}_R = M_\ell^{\rm diag}$~
and \,$U^\nu_L$\, ($N \times N$)\footnote{The case with $N>3$
contains additional $N-3$ light singlet sterile 
neutrinos and is awaiting confirmation from the 
ongoing MiniBooNE experiment at Fermilab\,\cite{MB}.} 
is from diagonalizing the neutrino mass matrix
(of Majorana or Dirac type)\footnote{The nonzero neutrino
masses can arise from a 
seesaw mechanism\,\cite{nu-seesaw,nu-seesaw1},
or a radiative mechanism\,\cite{nu-rad,nu-rad1}, or may have
a dynamical origin\,\cite{nu-dy,nu-dy1}.}.
Hence, the large or maximal mixings in the MNS matrix $V$
can originate from (i) either  $U^\nu_L$ (neutrino mass diagonalization),
(ii) or $U^{\ell}_L$ (left-handed lepton mass diagonalization), 
(iii) or both sources.
In the cases (ii) and (iii), we see that large mixings in $U^{\ell}_L$ 
play important roles for neutrino oscillation phenomena. Furthermore,
the right-handed lepton rotation $U^{\ell}_R$ does {\it not} enter the
MNS matrix and is thus free from the constraint of oscillation
experiments. This means that even in the case (i) large lepton mixings 
can originate from $U^{\ell}_R$ though $U^{\ell}_L$ is constrained
to have only small mixings. 
Clearly, the neutrino oscillation data alone could not identify the
{\it origin} of the MNS mixings (involving only the
product of two left-handed rotations), i.e., whether
the large or maximal mixings in MNS matrix really originate from the
neutrino mass matrix or charged lepton mass matrix or both\,\footnote{
It was shown that the large $\mu-\tau$ mixing can naturally arise
in GUT models such as ${\mathrm SU(5)}$ \cite{Alt}.
Recent studies\,\cite{ellis} also explored the attractive possibility 
that the maximal atmospheric neutrino mixing angle originates from 
the $\tau-\mu$ mixing in the charged lepton sector while the large
solar neutrino mixing angle comes from the mixing of $\nu_e -\nu_\mu$
in the neutrino mass matrix.}.\,  
Therefore, to fully understand the
flavor dynamics in the lepton and neutrino sector, 
it is important to directly test 
lepton rotation matrices $U^{\ell}_L$ and $U^{\ell}_R$ in other 
lepton flavor-violation processes. 
The large or maximal MNS mixings  observed in
the neutrino oscillation experiments strongly motivate searches
for large lepton flavor-violating interactions originating
from $U^{\ell}_L$ and $U^{\ell}_R$\,.

The effective dimension-six $\tau-\mu$ operator
Eq.\,(\ref{eq:taumu6}) can arise from exchange of certain heavy particles
such as a  heavier neutral gauge boson ($Z'$), or a 
heavier Higgs scalar ($S^{0}$).
The underlying dynamics for such interactions fall into
two distinct classes:
{\bf (a).} {\it flavor universal}, or,  
{\bf (b).} {\it flavor non-universal.} 
For the Class-(a), we can write a generic
lepton-bilinear term $\ov{\psi}\,\GA_j\,\psi$, where  
$\psi= \(e,\,\mu,\,\tau\)^T$. We thus deduce
\beq
\label{eq:bif}
\ov{\psi}\,\GA_j\,\psi =\left\{
\ba{l}
\dis\sum_{\chi(\neq\zeta)}\ov{\psi_\chi}\,\GA_j\,\psi_\zeta\,, 
~~~~~(\GA_j=1,\,\gamma_5 );
\\[6mm]
\dis\sum_{\chi}
\ov{\psi_\chi}\,\GA_j\,\psi_\chi\,, 
~~~~~(\GA_j=\gamma_\sigma,\,\gamma_\sigma\gamma_5);
\ea
\right.
\eeq
where $\chi,\zeta = L,R$.  The lepton mass diagonalization
$U^{\ell\,\dag}_L M_\ell U^{\ell}_R = M_\ell^{\rm diag}$
enables us to transform leptons  from flavor-eigenbasis
($\psi$) into mass-eigenbasis ($\psi'$) via
$\psi_L = U^\ell_L \psi_L'$  and
$\psi_R = U^\ell_R \psi_R'$.  
Thus, Eq.\,(\ref{eq:bif}) can be rewritten as, in the
mass-eigenbasis,
\beq
\label{eq:bim}
\ov{\psi}\,\GA_j\,\psi =\left\{
\ba{l}
\dis\sum_{\chi(\neq\zeta)}
\ov{\psi_\chi'}\,\GA_j\(U_\chi^{\ell\,\dag}U_\zeta^\ell\)\psi_\zeta'
~\To~ \O(1)\,\ov{\tau'}\GA_j\,\mu'
+ \O(1)\,\ov{\mu'}\GA_j\tau',
~~~~(\GA_j=1,\,\gamma_5 )\,;
\\[6mm]
\dis\sum_{\chi}
\ov{\psi_\chi'}\,\GA_j\,\psi_\chi'\,, 
~~~~(\GA_j=\gamma_\sigma,\,\gamma_\sigma\gamma_5) \,.
\ea
\right.
\eeq
Here, for $\GA_j = 1,\,\gamma_5 $,
the unsuppressed $\tau'-\mu'$ mixing bilinears can be induced
by the large  $\O(1)$  ``23'' or ``32'' entries in the lepton
rotation matrices 
$U_L^{\ell\,\dag}U_R^\ell$ and $U_R^{\ell\,\dag}U_L^\ell$.\,
We see that such large flavor-mixings between
$\tau'$ and $\mu'$ occur only for $\GA=1,\,\gamma_5$, implying
that for Class-(a) the $\tau-\mu$ operators 
arise from exchanging a 
heavy scalar (pseudoscalar)  in the underlying theory
(cf. Fig.\,2b and 2c). 
For scalar (pseudoscalar) type couplings to be flavor universal,
there should be certain flavor symmetry associated with
these couplings in the underlying theory. 
The quark bilinear $\ov{Q}\GA_jQ$ can join the same type
of flavor universal interactions, where $Q={\cal U},\,{\cal D}$
with ${\cal U}=(u,\,c,\,t)^T$ and ${\cal D}=(d,\,s,\,b)^T$.
But, without fine-tuning the flavor-violations in the quark-bilinear
would be relatively small
based on the experimental knowledge about the CKM matrix, except that
some right-handed mixings may be quite sizable.

We then proceed to consider Class-(b) with flavor non-universal
dynamics which is strongly motivated by the observed large 
mass hierarchies for leptons and quarks.
For instance, the induced lepton-bilinear may contain only the
third family tau-leptons 
as happened in the dynamical symmetry breaking models 
\cite{Hill-Rept,Hill-tc,Chiv} and
lepton non-universality models \cite{Ma,EL,Zhang}. 
Generically, we can write down a tau-lepton bilinear in the flavor
eigenbasis,
\,$
\ov{\tau_\chi}\,\GA_j\,\tau_\zeta 
$\,
where $\chi = \zeta$ corresponds to gauge boson $Z'$ exchange and 
$\chi \neq \zeta$ corresponds to scalar $S^0$ exchange.
The $Z'$ exchange from the strongly interacting theories such
as top-color models \cite{Hill-tc} is particularly interesting
for studying the effective dimension-6 operators (\ref{eq:taumu6})
as the $Z'$ induces strong couplings for Eq.\,(\ref{eq:taumu6}) 
which naturally fit the counting in Eq.\,(\ref{eq:C-def}). 
Because of the flavor non-unversality
of such underlying dynamics, the generic tau-lepton bilinear
takes the following form after the lepton mass-diagonalization,
\beq
\ov{\tau_\chi}\,\GA_j\,\tau_\zeta =
\ov{\ell_\chi^{\prime\a} }\,\GA_j\, 
( U_\chi^{\ell\,\ast})_{3\a}
( U_\zeta^\ell )_{3\b} \,\ell_\zeta^{\prime\b}
~\To~ \O(1) \,\ov{\tau'}\,\GA_j\,\mu' +
\O(1)\, \ov{\mu'}\,\GA_j\,\tau' 
\,,
\eeq
where $\chi\neq\zeta$ for $\GA_j=1,\,\gamma^{~}_5$ and
$\chi=\zeta$ for $\GA_j=\gamma^{~}_\sigma,\,
\gamma^{~}_\sigma\gamma^{~}_5$.
We see that, due to the allowed large entry
$(U_L^\ell)_{32}$ or $(U_R^\ell)_{32}$,
the un-suppressed flavor-violating $\tau-\mu$ bilinear 
can be generated for $\GA_j$ being either (axial-)vector or
(pseudo-)scalar (cf. Fig.\,2a-c), 
unlike the situation in Eq.\,(\ref{eq:bim}).
In most cases,
such flavor non-universal dynamics generically invokes the third family
quark-bilinears $\ov{t_\chi}\,\GA_j\,t_\zeta$  and 
$\ov{b_\chi}\,\GA_j\,b_\zeta$ at the same time, and some sizable
right-handed mixings 
such as $t_R^{~}\!-\!c_R^{~}$ mixing 
(or $b_R^{~}\!-\!s_R^{~}$ mixing)  
can naturally arise \cite{HY}.
To be model-independent, we will include all possible flavor
combinations in the quark-bilinear $\ov{q}^\a\,\GA_j\,q^\b$
[cf. Eq.\,(\ref{eq:taumu6})] for our phenomenological analysis in 
the next section.

\begin{figure}[H]
\vspace*{-10mm}
\begin{center} 
\hspace*{-15mm}
\includegraphics[width=20cm,height=6cm]{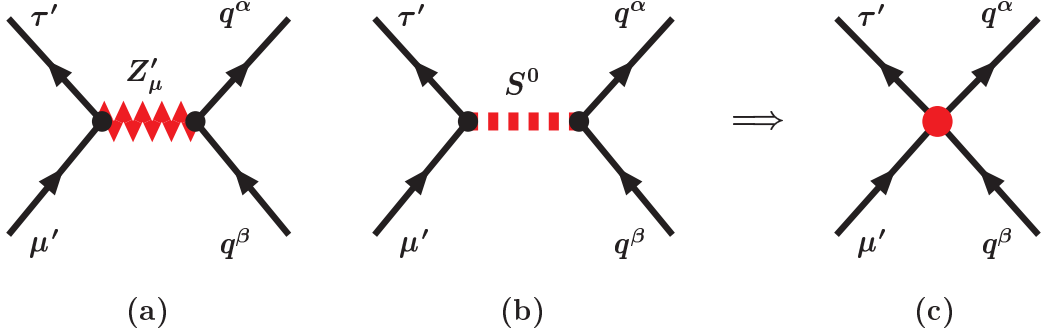} 
\vspace*{-10mm}
\caption{Illustration of typical generations of the 
effective four-Fermi interactions
[cf. Eq.\,(\ref{eq:taumu6})] in plot (c) from the exchange of
(a) a heavy new gauge boson $Z_\mu'$, or, 
(b) a heavy new scalar or pseudoscalar $S^0$ 
in the underlying theories. Here the $\tau'-\mu'$ flavor-violation
couplings solely come from the charged lepton mass-diagonalization
matrices $U^\ell_L$ and/or $U^\ell_R$.}
\end{center}
\end{figure}

In summary, a {\it full} understanding of leptonic flavor dynamics 
(for both masses and mixings) requires
experimental exploration not only via neutrino oscillations
but also via other lepton-flavor-violating processes.
The above classification shows that, 
given the large lepton mixings (especially for the tau and muon leptons)
as motivated by the atmospheric neutrino oscillation data, the effective
dimension-six flavor-violating $\tau-\mu$ 
operators (\ref{eq:taumu6}) can
be naturally realized without additional suppression (in contrast
to the quark sector).  Furthermore,
particular chirality structures (characterized by $\GA_j$)
can be singled out, depending on whether the underlying leptonic dynamics
is universal or non-universal in the flavor space.
The systematic phenomenology constraints analyzed in the
following Sec.\,III will demonstrate how the various low energy
precision data on lepton flavor-violations can probe the
new physics scale associated with the underlying flavor
dynamics, {\it complementary} to the neutrino oscillation
experiments.

\subsection{\large Radiative Corrections versus Leading Logarithmic Term}

In our analysis, we will consider two classes of bounds, 
from contributions at either the tree level or the loop level.
Many of these bounds arise from the tree-level operators
(\ref{eq:taumu6}) directly and will be derived from  
various low energy decay channels.  
In some cases, the
significant bounds can only be obtained by relating 
the operators involving one set of heavy quarks to those 
involving lighter quarks, through exchange of a $W$ or charged
Goldstone boson at the loop level (cf., Fig.\,3 in Section 3).   
How does one handle radiative loop effects in an effective
theory? For such calculations, typically, some loop 
integrals are divergent and must be cut off at a scale
$\Lambda$.   In the case of $W$ or Goldstone boson exchange, 
the divergence is logarithmic.  
We perform the loop integral 
by retaining the leading logarithmic terms 
in which the ultraviolet (UV) cutoff
$\cut$ would reliably represent the scale 
of new physics\,\cite{BL}.
In doing the analysis, we assume only one 
$(\bar{\mu}\Gamma_j\tau)(\bar{q}^{\a}\Gamma_jq^{\b})$ operator
to be  non-zero\footnote{If 
we were to keep other tree-level light-quark operators as well, 
then the loop contributions (induced by the heavy quark operators)
will essentially renormalize the tree-level 
light-quark operators.   In that case, we could make a 
{\it naturalness assumption}---that there is no 
accidental cancellation between the
renormalized tree-level coefficient of a given
light-quark operator and the leading logarithmic contributions from
the corresponding heavy-quark operator (cf. Fig.\,3).  We can
thus estimate the renormalized effective 
light-quark operator coupling 
by using the leading logarithmic terms, resulting from
cutting off the divergent integrals 
under the $\ov{\rm MS}$ scheme. The renormalization group running of the
coefficient from the scale $\mu=\cut$ 
down to the relevant low energy scale 
$\mu =\mu_0$ induces the leading logarithmic contribution of 
$\ln (\Lambda/\mu_0)$ which may be used for estimating the size of
radiative loop corrections.}.
Other diagrams involving closed fermion loops may have 
quadratic divergences.  Unlike the logarithmic terms, such $\O(\cut^2)$ power 
corrections are not guaranteed to always represent the 
real contributions of the lowest heavy physical state 
of a mass $M_{\rm phys}$, so that to be conservative it is usually 
suggested\,\cite{BL} 
that one only uses the logarithmic terms ($\sim\ln \cut$) computed in
the effective theory, for representing the new physics 
contribution ($\sim\ln M_{\rm phys}$) from an underlying full theory.
We will take this approach for the loop analysis in Section 3,
though we keep
in mind that retaining only leading logarithms may possibly 
underestimate the new physics loop-contributions if the
$\O(M_{\rm phys}^2)$ terms are not vanishing in a given
underlying theory. This exception occurs only when the heavy
mass effect in the underlying theory does not obey the usual
decoupling theorem \cite{DCT}\footnote{One typical example is the
models with heavy chiral fermions. In the case of a heavy SM Higgs,
the $\O(M_H^2)$ power corrections show up only at two-loop level
due to the screening theorem\,\cite{scr}, but an extended Higgs sector
may possibly escape the screening theorem at one-loop. 
In general, for any heavy state
of mass $M_{\rm phys}$, the nondecoupling occurs so long as 
$M_{\rm phys}$ is proportional to certain coupling of this state with
light fields (which remain in the low energy theory).}.  
Thus, extracting the
possible nonzero $\O(M_{\rm phys}^2)$ terms is a highly model-dependent
issue and is hard to generally handle in an 
effective theory formalism. The traditional ``leading logarithm''
approach provides a conservative estimate for the effective theory
analysis and is justified for those underlying theories in which
the effects of the heavy states (integrated out from the
low energy spectrum) exhibit the decoupling behavior.

\section{Phenomenological Constraints}

We consider the general $\tau-\mu$ operators in Eq.\,(\ref{eq:taumu6}) 
and take $\GA_j$ to
be the same for  both the lepton and quark pieces. There are four types of
operators to be considered, $\GA_j = (S,P,V, A)$, 
and for each type there are twelve combinations of $q_\alpha q_\beta$,
$(uu,dd,ss,cc,bb,tt,ds,db,sb,uc,ut,ct)$. 
This gives a  total of 48 operators for our analysis.  We first consider
operators involving two light quarks ($uu,dd,ss,ds$), 
then the non-diagonal operators involving one or more
heavy quarks ($db,sb,uc,ut,ct$), and finally consider 
the diagonal operators involving two heavy quarks ($cc,bb,tt$).
In our analysis, we will consider one operator to be nonzero at a time
and derive the corresponding bound on the new physics scale $\cut$.
This should provide a sensible estimate of the scale $\cut$  under the
{\it naturalness assumption} mentioned in Section\,II,  
which states that there is no accidental cancellation 
among the contributions of different operators.

\subsection{\large Operators with Two Light Quarks}

For operators with two light quarks, 
the neutrinoless decay of the $\tau$ into a $\mu$ and one or more light
mesons will provide the best bound.   First, we establish our conventions:
The PCAC condition for the pseudoscalar octet gives  
\begin{equation}
\left< 0 \left| j_\mu^{(5)b}(0) \right| \phi^a (p) \right> 
= i \delta^{ab} \frac{F_\phi}{\sqrt
2} p_\mu \,,
\end{equation} 
where $F_\phi$ is the meson decay constant and the Cartesian 
components of the axial vector current are
\begin{equation} {j^\mu}^{(5)b} = \bar q \gamma^\mu \gamma^5 
\frac{\lambda^b}{2} q \,.
\end{equation} Here $\phi = \phi^a \lambda^a / {\sqrt 2}$ 
with $a = 1 \cdots 8$ is the $3 \times 3$ matrix of 
pseudoscalar meson fields and $\lambda^a$ are the Gell-Mann 
matrices normalized according to ${\rm Tr} (\lambda^a
\lambda^b ) = 2 \delta^{ab}$.   
For current quark masses, we choose 
\begin{equation}
m_u=m_d=5\ \mev \,,\quad {\rm and} \quad m_s=120\ \mev \,,
\end{equation}
and the results are not particularly sensitive to these choices.  
For simplicity, the muon mass (and the pion
mass, when applicable) will be neglected relative to the tau 
mass when calculating kinematics.

Knowing the vacuum transition matrix element, we can readily 
evaluate the particle decay width. For instance, for a two-body
decay $\tau\to \mu M$ where $M$ is a generic light meson, we
have the spin-summed and averaged partial width
\begin{equation}
\Gamma(\tau\to \mu M) = {1\over 2m_\tau} {1\over (8\pi)}
\overline{\sum} |{\cal M}|^2,
\end{equation}
where the transition amplitude $i{\cal M}=<\mu M|i{\cal H}|\tau>$
can be evaluated by vacuum insertion.

\vskip 0.4 cm
\noindent
\underline{\bf {\Large $\bullet$} 
Axial Vector Operators}
\vspace*{3mm}

Bounds on the $uu$ and $dd$ axial operators can be obtained by looking 
at $\tau\rightarrow \mu\pi^0$. Using 
\begin{equation}
\left< 0 \left|  \bar u \gamma^\mu \gamma^5 u \right| \pi^0 (p) 
\right> = \left< 0 \left| \bar q \gamma^\mu
\gamma^5\left( \frac {1}{2} + \frac{\lambda^3}{2} \right)  q 
\right| \phi^3 (p) \right> =  i
\frac{F_\pi}{\sqrt 2} p_\mu 
\end{equation} 
with $F_\pi = 131\ {\rm MeV}$,
and noting that for $u \rightarrow d$ the right-hand side is the 
same except with an opposite sign, we find 
\begin{equation}
\Gamma \left( \tau \rightarrow \mu \pi^0 \right) = 
\frac{1}{\Lambda^4}\frac{\pi}{2} F_\pi^2 m_\tau^2 < 0.908
\times 10^{-17} ~ {\rm GeV} \,,
\end{equation} 
where the inequality comes from the  $90\%$\,C.L.
experimental bound on this decay mode listed in Ref.\,\cite{pdg}.
This then implies that, for both the 
~$\bar u \gamma^\mu \gamma^5 u$~ and 
~$\bar d \gamma^\mu \gamma^5 d$~ operators,  
\begin{equation}
\Lambda > 11.3 ~{\rm TeV}\,. 
\end{equation}

For the $ss$ axial operator, a bound is obtained from 
$\tau\rightarrow \mu\eta$.   We have 
\begin{equation}
\left< 0 \left|  \bar s \gamma^\mu \gamma^5 s \right| 
\eta (p) \right> = i \frac{-2}{\sqrt 3}
\frac{F_\eta}{\sqrt{2}}\, p^\mu  \,,
\end{equation} 
where 
$\dis F_\eta = F_\eta^8 - \dis\frac{1}{\sqrt{2}}F_\eta^0$ 
is defined using $\left< 0 \left| \bar q \gamma^\mu \gamma^5  
\frac{\lambda^{8,0}}{2}  q  \right| \eta (p) \right> =  
i \frac{F_\eta^{8,0}}{\sqrt 2} p_\mu$.  
Here, \,$\lambda^0 =\dis {\sqrt{\frac{2}{3}}}\,${\bf 1}\, and at 
the next-to-leading order (NLO) of the chiral 
perturbation theory \cite{feldman}, 
$F_\eta^8 = 154 \ {\rm MeV}$ and $F_\eta^0 = 25\ {\rm MeV}$.  
Using these values we find that $
F_\eta \approx F_\pi$, the value from the ${\mathrm SU(3)}$ limit.   
This gives,
\begin{equation}
\Gamma(\tau \rightarrow \mu \eta) = \frac{ 2\pi}{3} 
\frac { {\left( m_\tau^2 - m_\eta^2\right )}^2}{m_\tau}
\frac{F_\eta^2}{\Lambda^4} < 2.179 \times 10^{-17} ~ {\rm GeV} \,,
\end{equation} 
implying
\begin{equation}
\Lambda > 9.5  ~{\rm TeV}.
\end{equation}
Note that, for the isospin-invariant effective operator 
$\dis\f{4\pi}{\Lambda^2} \left( \bar \mu A \tau \right) 
\left( \bar u A u + \bar d A d  \right)$ with $A = \gamma^\mu
\gamma^5$, the same bound of $\cut > 9.5$\,TeV 
can be derived from the above process.

For the $sd$ operator, the bound comes from $\tau\rightarrow \mu K^0$.  
Thus, we have
\begin{equation}
\left< 0 \left| \bar s 
\gamma^\mu \gamma^5 d \right| K^0 \right> 
= i F_K p^\mu \,,
\end{equation} 
where experimentally 
$F_K = 160\ {\rm MeV}$.  
This leads to
\begin{equation}
\Gamma(\tau \rightarrow \mu K^0)=  \pi  \frac { {\left( m_\tau^2 - 
m_K^2\right )}^2}{m_\tau}
\frac{F_K^2}{\Lambda^4} < 2.27 \times 10^{-15} ~ {\rm GeV} \,,
\end{equation} 
and thus,
\begin{equation}
\Lambda > 3.6 ~ {\rm TeV}.
\end{equation}

\vskip 0.4cm
\noindent
\underline{\bf {\Large $\bullet$} 
Pseudoscalar Operators}
\vspace*{3mm}

Here, the Dirac equation is used to reduce the 
axial vector matrix elements to pseudoscalar matrix elements,
and then we use the same processes as above. 

We find that
\begin{equation}
 \left< 0 \left| \bar u \gamma^5 u \right| \pi^0 (p) \right> 
= -  \left< 0 \left| \bar d \gamma^5 d \right|
\pi^0 (p) \right> 
= \frac {i}{\sqrt 2} \frac{ m_\pi^2 }{~m_u +m_d~} F_\pi \,,
\end{equation} 
which then yields
\begin{equation}
\Gamma (\tau \rightarrow \mu \pi^0) 
= \frac{\pi}{8} \frac{ F_\pi^2 m_\pi^4 m_\tau}{\Lambda^4 m_q^2} < 0.908
\times 10^{-17} ~ {\rm GeV} \,,
\end{equation}
so that
\begin{equation}
\Lambda > 11.7 ~ {\rm TeV}.
\end{equation}

For the strange quark operator, we find
\begin{equation}
 \left< 0 \left| \bar s \gamma^5 s 
\right| \eta_8 (p) \right> 
= - i {\sqrt 6} \,F_\eta^8 
\f{m_{\eta_8}^2}{~m_u + m_d + 4 m_s~}
\end{equation} which gives (taking $\eta = \eta_8$)
\begin{equation}
\Gamma (\tau \rightarrow \mu \eta) = \frac{6 \pi}{\Lambda^4} \frac{m_\eta^4 {F_\eta^8}^2}{ {\left( m_u + m_d +
4 m_s \right)}^2} 
\frac{ {\left( m_\tau^2 - m_\eta^2 \right) }^2}{m_\tau^3} 
< 2.18 \times 10^{-17} ~ {\rm GeV} \,,
\end{equation} 
implying
\begin{equation}
\Lambda > 9.9 ~ {\rm TeV}.
\end{equation}
For the isospin-invariant effective operator 
$\dis\frac {4 \pi} {\Lambda^2} \left( \bar \mu P \tau \right) 
\left( \bar u P u + \bar d P d  \right)$ with $\,P = \gamma^5$,\, 
the same bound of $\cut > 9.9$\,TeV 
can be derived from the above process.

Finally, we have 
\begin{equation}
\left< 0 \left| \bar s \gamma^5 d \right| K^0 \right> = 
i \frac{m_{K^0}^2}{m_d + m_s} F_K  \,,
\end{equation} 
which gives
\begin{equation}
\Gamma(\tau \rightarrow \mu K^0) =  
\frac{\pi}{\Lambda^4} \frac{m_K^4 {F_K}^2}{ {\left( m_d +  m_s \right)}^2}
\frac { {\left( m_\tau^2 - m_K^2 \right) }^2}{m_\tau^3} 
< 2.27 \times 10^{-15} ~ {\rm GeV} \,.
\end{equation} 
Thus, we deduce
\begin{equation}
\Lambda > 3.7 ~ {\rm TeV}.
\end{equation}

\vskip 0.4cm
\noindent
\underline{\bf {\Large $\bullet$} 
Vector Operators}
\vspace*{3mm}

We take a simple ${\mathrm SU(3)}$  relation:
\begin{equation}
\left< 0 \left| \bar q \gamma^\mu \frac{\lambda^a}{2} 
q \right| V^b \right> = i \epsilon^\mu \delta^{ab}
\frac{c}{\sqrt 2} \,, 
\end{equation} 
where $V_\mu =\dis V_\mu^a \frac{\lambda^a}{\sqrt 2}$ 
is the vector meson octet.  
Using vector meson dominance \cite{sakurai}, 
we determine the dimensionless ratio
$g\equiv m^2/c$ from $V\to e^+e^-$ for each vector 
$V=\rho, \omega, \phi$, which
yields  $g_\rho = 5.1$, $g_\omega = 17$ and $g_\phi = 12.9$.  
These phenomenological values indicate some ${\mathrm SU(3)}$(3) breaking, 
as expected.  Assuming ideal mixing, we get
\begin{eqnarray}
\left< 0 \left| \bar u  \gamma^\mu u \right| \rho^0 \right> 
& = & - \left< 0 \left| \bar d \gamma^\mu d \right|
\rho^0 \right> = i \epsilon^\mu K_\rho \,, 
\nonumber \\ 
\left< 0 \left| \bar s  \gamma^\mu s \right| \phi \right> 
& = & i \epsilon^\mu K_\phi \,, 
\nonumber \\ 
\left< 0 \left| \bar s  \gamma^\mu d \right| {K^0}^*  \right> 
& = & i \epsilon^\mu K_{K^*} \,,
\end{eqnarray} 
where
\begin{equation} K_\rho = \frac{m_\rho^2}{g_\rho} \,, 
~ \quad K_\phi = \frac{3 m_\phi^2}{g_\phi} \,,
\end{equation} 
and in the ${\mathrm SU(3)}$ 
limit, one has \,$K_{K^*}=K_\rho$\,. 
Then, we find
\begin{equation}
\Gamma(\tau \rightarrow \mu V) = 
\frac{\pi K_V^2}{\Lambda^4} \frac{m_\tau}{m_V^2} \left[ m_\tau^2 {\left( 1 -
\frac{m_V^2}{m_\tau^2} \right)}^2 {\left( 1 + 2  \frac{m_V^2}{m_\tau^2} 
\right)}  - 3 \frac{m_\mu m_V^2}{m_\tau^3} \right] ,
\end{equation} 
which gives bounds as follows:
\beq
\ba{rclclc}
\Gamma(\tau \rightarrow \mu \rho) &<& 1.43  
\times 10^ {-17} ~{\rm GeV} &~~\Rightarrow~~ & \Lambda > 12.4 ~{\rm
TeV}, ~&~ ({ uu,dd}) \,, 
\nonumber \\
\Gamma(\tau \rightarrow \mu \phi) &<&  1.59 
\times 10^ {-17}~ {\rm GeV} 
&~~\Rightarrow~~ & \Lambda > 14.3 ~  
{\rm TeV},  ~&~  ({ ss}) \,,
\nonumber \\
\Gamma(\tau \rightarrow \mu K^*) &<& 1.7 \times 10^ {-17} 
~{\rm GeV} &~~\Rightarrow~~ & \Lambda > 12.8~ 
{\rm TeV},   ~&~ ({ds}) \,.
\ea
\eeq

\vskip 0.4cm
\noindent
\underline{\bf {\Large $\bullet$} 
Scalar Operators}               
\vspace*{3mm}

Scalar operators will lead to three-body decays of the 
$\tau$ into a $\mu$ and two mesons.  Using the leading
order chiral Lagrangian, we obtain the matrix elements of scalar
densities at the origin
\begin{eqnarray} 
\la 0|\bar s s |K^+ K^- \ra = \la 0 | \bar u u
| K^+ K^- \ra = B_0 \,,
\nonumber \\ 
\la 0 | \bar u u | \pi \pi \ra  = \la 0 | \bar d d | \pi \pi \ra = 
B_0 \,, 
\nonumber \\
 3 \la 0 | \bar u u | \eta_8 \eta_8 \ra 
= \frac{3}{4} \la 0 | \bar s s | 
\eta_8 \eta_8 \ra = B_0 \,,
\nonumber \\
 \la 0 | \bar d s | \pi^+ K^- \ra = B_0 \,,
\end{eqnarray} 
where $m_\pi^2 = 2 \hat m B_0$ and $m_{K^+}^2 = (m_u + m_s) B_0$ 
with $\hat m = m_u + m_d$.  We take 
$m_u = m_d = 5$\,MeV, 
which gives $B_0 = 1.96$ GeV and $m_s = 120$\,MeV.  
Thus, the differential decay widths are computed as
\begin{eqnarray} 
d\Gamma (\tau \rightarrow \mu \pi^0 \pi^0 ) 
&=& \frac{1}{2}d\Gamma (\tau \rightarrow \mu \pi^+
\pi^- ) = \frac{B_0^2}{64 \pi^3 {\Lambda_{uu,dd}}^4} ( E_1 +
m_\mu ) dE_1 dE_2 \,,\nonumber 
\\[1mm] 
d\Gamma (\tau
\rightarrow \mu \eta \eta ) &=& 
\({\frac{B_0}{3}}\)^2\frac{1}{64 
\pi^3 \Lambda_{uu,dd}^4} ( E_1 + m_\mu ) dE_1
dE_2  \,,
\nonumber 
\\[1mm] 
&=& \({\frac{4 B_0}{3}}\)^2\frac{1}{64 \pi^3 \Lambda_{ss}^4} 
( E_1 + m_\mu ) dE_1 dE_2  \,,
\nonumber  
\\ [1mm]
d\Gamma (\tau \rightarrow \mu K^+ K^-) 
&=& {B_0}^2\frac{1}{32 \pi^3 \Lambda_{dd,ss}^4} (
E_1 + m_\mu ) dE_1 dE_2 \,,
\nonumber 
\\[1mm] 
d\Gamma (\tau \rightarrow \mu K^+ \pi^-) &=& {B_0}^2\frac{1}{32 \pi^3
\Lambda_{ds}^4} ( E_1 + m_\mu ) dE_1 dE_2 \,.
\end{eqnarray}

The best bounds on the $\bar{u} u$ and $\bar{d}d$ operators come from  
\begin{equation}
\Gamma(\tau \rightarrow \mu \pi^+ \pi^-) < 0.186 
\times 10^{-16} ~ {\rm GeV}, \quad \Rightarrow \quad
\Lambda_{uu,dd} > 2.6 ~ {\rm TeV},
\end{equation}
the bound on the $\bar{s}s$ operator from 
\begin{equation}
\Gamma(\tau \rightarrow \mu K^+K^-) < 0.341 \times 10^{-16} 
\quad {\rm GeV}, \quad  \Rightarrow \quad
\Lambda_{uu,ss} > 1.5 ~ {\rm TeV},  
\end{equation}
and the bound on the $\bar{s}d$ or $\bar{d}s$ operator from 
\begin{equation}
\Gamma(\tau \rightarrow \mu K^+ \pi^-) < 0.168 \times 10^{-16} 
~ {\rm GeV}, \quad \Rightarrow \quad
\Lambda_{ds} > 2.3 ~ {\rm TeV}.
\end{equation}

In summary of this subsection, we see that the bounds on operators 
involving two light quarks range from $1.5$ to $14.5$\,TeV.  Improvement in
the experimental limits on the branching ratios, of course, 
will increase these bounds via
the fourth root of the branching ratio.  
Much better improvement can be obtained
from processes involving decays of heavy quarks, 
which we consider below.

\subsection{\large Non-diagonal operators involving one or more heavy quark}

Now, we analyze the operators involving ($uc,ut,ct,db,sb$) quarks.   
The bounds involving an up quark and a charm
quark are problematic since the $D^0$ can not (barely) 
decay into $\mu\tau$ because of kinematics. 
The bounds on
these operators will be discussed at the end of this subsection.   
We first turn to the $B$ meson decays.

\vskip 0.4cm
\noindent
\underline{\bf {\Large $\bullet$} 
$\mbox{\boldmath $B$}$ meson decays}
\vspace*{3mm}

Using the techniques described in the above subsection, 
one can bound the pseudoscalar and axial vector
operators for $\bar{b}d$ quarks by looking at $B^0\rightarrow \mu\tau$.  
For the axial vector operator, we find
that, using the experimental limit of $8.3\times 10^{-4}$ 
on the branching ratio
\begin{equation}
\Gamma(B\rightarrow\mu\tau) 
= \frac{\pi f^2_B m_B m^2_\tau}{\Lambda^4} 
\left(1-\frac{m^2_\tau}{m^2_B}\right)^2,
\quad \Rightarrow \quad \Lambda > 8.2~ {\rm TeV},
\end{equation}
where we take $f_B = 200$\,MeV.  
For the pseudoscalar operator we find,
\begin{equation}
\Gamma(B\rightarrow\mu\tau) = \frac{8 m_B^2 \beta^3_b}{\sqrt{\pi}\Lambda^4}
\left(1-\frac{m^2_\tau}{m^2_B}\right)^2,
\quad \Rightarrow \quad \Lambda > 9.3~ {\rm TeV}.
\end{equation}
In this latter case, we have used the result from 
Sher and Yuan \cite{sher91}, 
$|\langle 0|\bar{d}\gamma^5 b|B\rangle|^2 = 4 m_B 
{\beta_B^3}/{\pi^{3/2}}$, 
where $\beta_B \approx 300 $\,MeV 
is a variational parameter.

In a moment, we will consider the scalar and vector operators, 
but let us first look at the pseudoscalar and
axial vector cases for $\bar{b}s$ quarks.  Here, precisely the same 
analysis as for $B\rightarrow \mu\tau$ can be
done for $B_s\rightarrow\mu\tau$, with the same result for the width, 
in the approximation
where the masses of the constituent $s$ and $d$ quarks are equal.  
Alas, there are no published experimental bounds for
$B_s\rightarrow\mu\tau$.   Note that the lifetime of the
$B_s$ is given by $1.46\pm 0.06$ picoseconds, compared with the $B$ 
lifetime of $1.54 \pm 0.02$ picoseconds.  
There are consistent, as expected.   But, if the rate for
$B_s\rightarrow \mu\tau$ were too large, then the lifetime 
would be substantially shorter.  A 10 percent
branching ratio would shorten the lifetime by about $0.14$ picoseconds, 
which would lead to a significant
discrepancy.  Without a detailed analysis, one can just conclude 
that there is a bound of five to ten percent on
the branching ratio for
$B_s\rightarrow\mu\tau$; we will give bounds assuming it is ten percent.

With a $10\%$ bound, the above results scale as the fourth root of 
the branching ratio, giving a bound on the
axial vector operator of $\Lambda > 2.5$\,TeV and on the pseudoscalar 
operator of $\Lambda > 2.8$\,TeV.

Note that here is a place where an experimental bound on 
$B_s\rightarrow\mu\tau$ would be very useful.  This
decay is particularly important because {\it all} of the quarks 
involved are second and third generation, and new
physics effects might be  substantial (especially if related 
to symmetry breaking); this decay also conserves
``generation" number, and is thus particularly interesting.

We now turn to the scalar and vector operators.  
Here the matrix elements $\langle K|
\bar{s}\gamma^\mu b|B\rangle$ and
$\langle \pi|\bar{d}\gamma^\mu b|B\rangle$ are needed, 
along with their scalar counterparts.
The vector matrix elements have been calculated in a 
quark model by Isgur, Scora, Grinstein
and Wise \cite{isgur89}.  They note that for a light 
pseudoscalar meson $X$,
\begin{equation}
\left< X(p_X)|\bar{q}\gamma^\mu b|\bar{B}(p_B)\right> 
\equiv f_+(q^2) (p_B+p_X)^\mu + f_-(q^2)
(p_B-p_X)^\mu  \,,
\label{matrixelements}
\end{equation}
and present expressions (in their Appendix\,B) for $f_+$ and $f_-$.   Here
$q^2=(p_B-p_X)^2$ and $B$ and $X$ are on-shell with $X\sim \bar{q}d$. 
The masses in these
expressions are constituent quark masses, which we take to be
$m_q=300$ MeV, and we also take their variational parameters,
$\beta$, to be $300$\,MeV. 
For instance,  these values give 
\begin{equation}
f_+(q^2) \approx -f_-(q^2) \approx
\frac{3\sqrt{2}}{8}\sqrt{\frac{m_b}{m_q}}
\exp\[\frac{m_X-E_X}{2\kappa^2 m_X}\] ,
\label{formfactor}
\end{equation}
where $\kappa\approx 0.7$ is a relativistic compensation and where 
$m_X\approx 2m_q$.
 As a result of these approximations, the matrix elements
should be taken {\it cum grano salis}, with an error that could be a 
factor of $2 - 4$
(which translates into a factor of $1.2 - 1.4$ uncertainty in the final 
results for $\Lambda$).  The uncertainty might be somewhat larger 
for the matrix elements involving
the pion, since the relativistic compensation factors are suspect.

The result for the vector couplings is given (illustrating the 
$B\rightarrow K\mu\tau$ case) by
\begin{equation}
\frac{d\Gamma}{dE_\mu dE_\tau}= \frac{|{\cal M}^2|}{~32\pi^3 m_B~} \,,
\end{equation}
where
\begin{equation}
\left|{\cal M}^2\right|=\frac{18\pi^2 m_B}{\Lambda^4 m_d}\ (A_1A_2-A_3)\ 
\exp\[-\dis\frac{m_B-E_\tau-E_\mu-m_K}{m_d}\] ,
\end{equation}
with 
$A_1=m^2_B-m^2_K-m^2_\tau-2m_BE_\mu,\, A_2=
m^2_B-m^2_K+m^2_\tau-2m_BE_\tau$,\, and $A_3=
m^2_K[-m^2_B-m^2_\tau+m^2_K+2(E_\mu+E_\tau)]$\,.

What are the experimental bounds for 
$B\rightarrow K\mu\tau$ and $B\rightarrow \pi\mu\tau$?  None are listed.
If the $\tau$ decays semi-hadronically (which occurs 65 percent 
of the time) then $B\rightarrow
K\mu\tau$ will look like $B\rightarrow X_c \mu\nu$ \cite{urheimfirst}.  
Then measurements of $B\rightarrow X_c
\mu\nu$ would give a higher rate than for $B\rightarrow X_c e \nu$.  
These have been
measured separately, with accuracies better than 0.5\%, and thus an excess of
$\mu$-like events have not been seen with a sensitivity of 
1.5\% at 90\% confidence level.  If one assumes that
the probability of classifying $B\rightarrow K\mu\tau$ decays 
as $B\rightarrow X_c\mu\nu$ is smaller by a
factor of two, then, folding in the 65\% branching fraction 
into hadrons, one would get a limit of 1.5\%
divided by $0.5$ for acceptance and $0.65$ for the branching fraction, 
which is about 5 percent.   A very
similar argument would apply to $B\rightarrow \pi\mu\tau$.  
Obviously, a more detailed analysis could yield a
substantially better bound.  However our result only scales as 
the fourth root of the branching ratio bound and,
with the relatively large uncertainty in the matrix elements, 
one probably can't do much better.
With a 5 percent branching ratio, we find that the bound on 
the $(bs)$ vector operator is 2.6 TeV,
and the bound on the $(bd)$ vector operator is 2.2 TeV.

For the scalar operator, one must differentiate the vector 
matrix elements in Eq.\,(\ref{matrixelements}), 
taking care to properly include the
$\exp\[-i(p_B-p_K)\cdot x\]$ 
factors.  We find that, for instance,
\begin{equation}
\left< K(p_K)|\bar{s}b|\bar{B}(p_B)\right> \approx \frac{1}{m_b} 
\left[ f_+(q^2) (m_B^2-m^2_K)
+ f_- (q^2)(m^2_B+m^2_K-2m_BE_K)\right] ,
\end{equation}
where $f_\pm (q^2)$ are given in Eq. (\ref{formfactor}).
Using this matrix element, we find the bound on the $(bs)$ 
scalar operator is $2.6$ TeV, and
similarly that for the $(bd)$ scalar operator is $2.2$ TeV.   
One should keep in mind the
relatively large uncertainties in these bounds due to the 
hadronic uncertainties discussed above.

\vskip 0.4cm
\noindent
\underline{\bf {\Large $\bullet$} 
Top Quark decays }
\vspace*{3mm}

Due to the very short life-time of a top quark,
bound state top mesons do not exist. Bounds on operators
involving a top quark can be readily obtained by looking for 
 decays $t\rightarrow c\mu\tau$ and $t\rightarrow u\mu\tau$.
Neglecting the final state masses, one finds that the width is
\begin{eqnarray}
\Gamma = \left\{ \begin{array}{ll}
\dis {{m_t^5}\over {96\pi\Lambda^4}}, & 
 \mbox{for scalar and pseudo scalar couplings,}\\[5mm]
\dis \frac{m_t^5}{24\pi\Lambda^4},    & 
  \mbox{for vector and axial vector couplings.}
                    \end{array}
\right.
\end{eqnarray}
CDF \cite{CDF} measures the ratio, 
\begin{equation} 
{\mathcal R} 
= \dis\f{~{\rm Br}(t\rightarrow Wb)~}{{\rm Br}(t\rightarrow Wq)} \,, 
\end{equation}
by counting $b$-tagged top events and all top to $Wq$ events.  
The result is $\R = 1.23^{+ 0.37}_{-0.31}$,
which translates into $\R > 0.72$ at $90\%$ C.L.
For the $t\rightarrow u\mu\tau$
channel, one considers this to be similar to $Wq$, in that the
signature is an isolated muon with some jet activity (clearly, a
detailed analysis by CDF/D0 could distinguish between $Wq$ and
$u\mu\tau$),  
and thus one approximately has Br$(t\rightarrow u\mu\tau) <
1-0.72 = 0.28$ at one standard deviation. This leads to a constraint 
\begin{eqnarray}
\Lambda > \left\{ \begin{array}{ll}
190 \ \gev , & \mbox{for scalar and pseudoscalar couplings,}
\\[2mm]
270 \ \gev , & \mbox{for vector and axial vector couplings.}
                    \end{array}
\right.
\end{eqnarray}

It should be kept in mind that the above bounds on $\cut$
are so close to the top quark mass that
the use of the effective field theory is not reliable.  
As discussed in Sec.\,IIB, for models obeying the
Naive Dimensional Analysis (NDA), the corresponding
bounds become stronger than the above by about
a factor of $\sqrt{4\pi}\simeq 3.5$, and thus in this case the application
of effective theory formalism will be more reasonable.  
One could improve on the above bounds significantly from non-observation of 
\,$t\rightarrow {\rm jet}+\mu\tau$\, decay, 
but this has yet to be done.  (We also recall that our final limits 
on $\Lambda$ only vary as the fourth root of the branching ratio.)

\vskip 0.4cm
\noindent
\underline{\bf {\Large $\bullet$} 
Loop Contributions}
\vspace*{3mm}

Operators involving heavy quarks are harder to constrain with
meson decays. However, one-loop contributions via $W^\pm$ exchange
may mediate the transition from a heavy quark to a lighter one.
This leads to processes with external light quarks and thus results
in possibly significant constraints from light mesons.
For the vector and axial vector couplings, strong bounds 
can be obtained by considering the loop contributions
with $W^\pm$ exchange. Such contributions will also give good bounds 
on the $(cu),\,(cc),\,(tt)$ operators, as will be discussed below.

\begin{figure}
\begin{center}
\epsfysize 2.2in \epsfbox{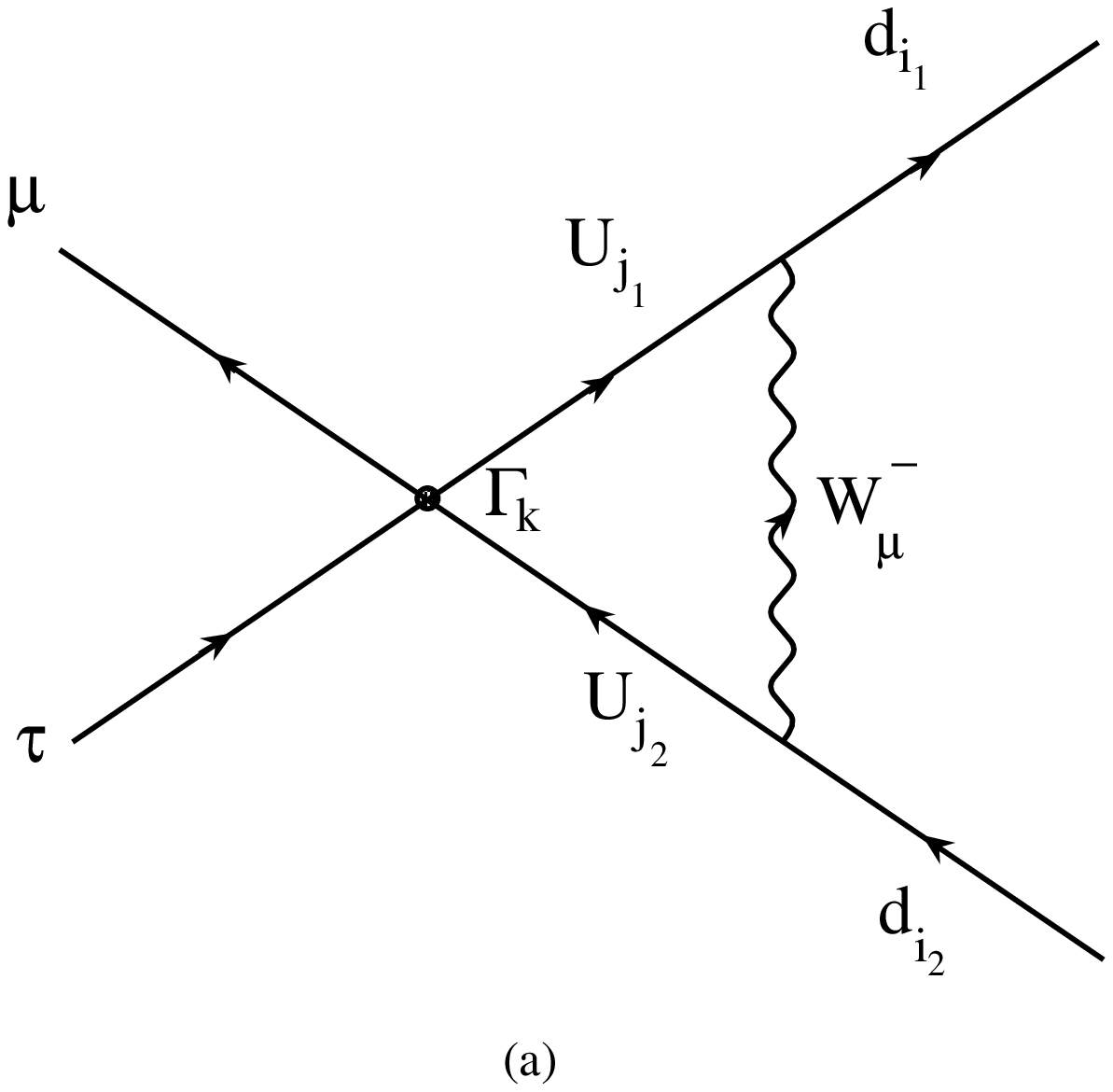}
\hspace*{8mm}
\epsfysize 2.2in \epsfbox{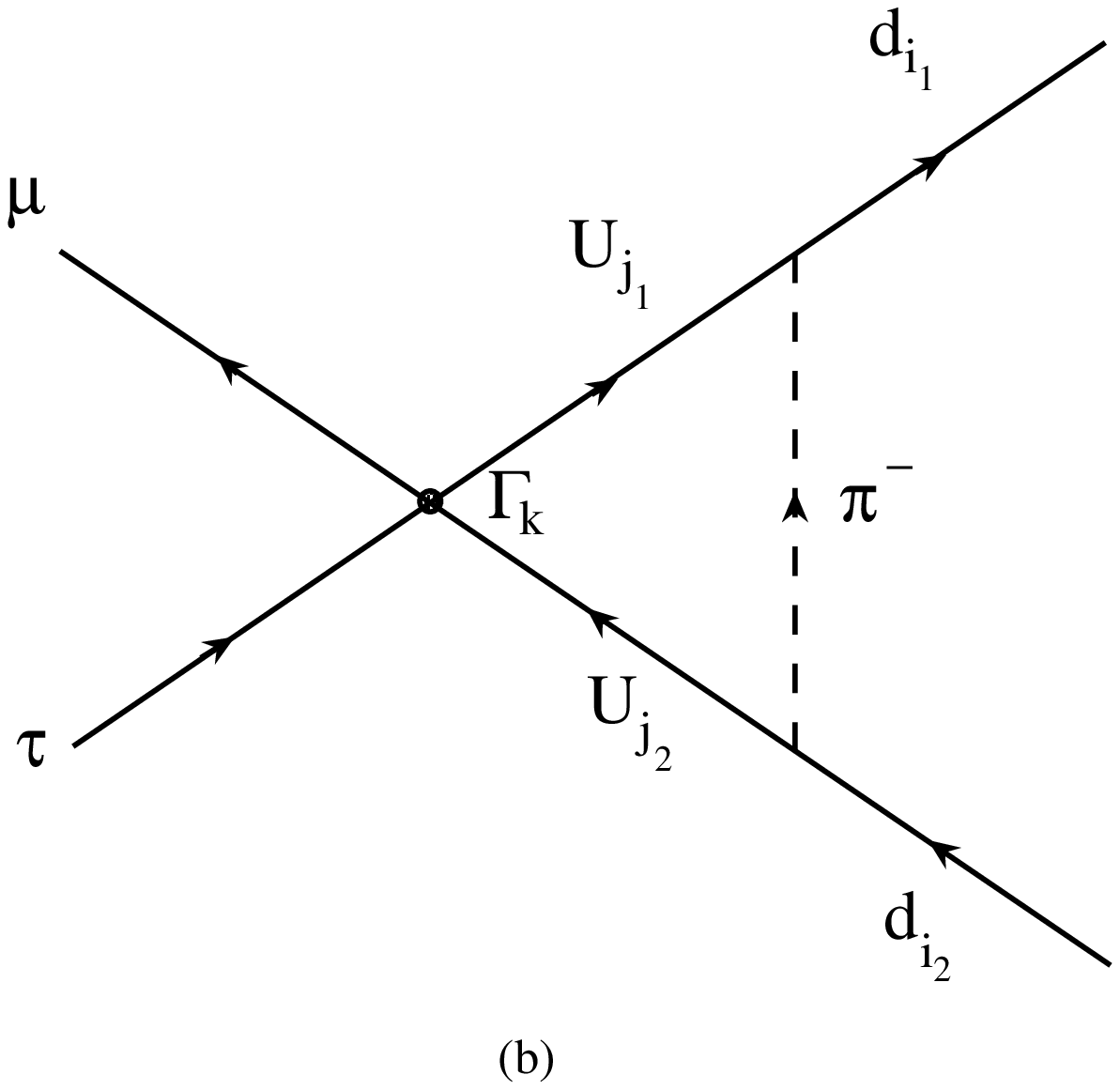}
\caption{Diagrams which relate couplings involving $u$-type quark 
bilinears to $d$-type quark bilinears.   
The $\pi^-$ is the charged Goldstone boson.}
\end{center}
\end{figure}

Consider the loops in Fig.\,3(a) and (b), 
where $\pi^-$ is the charged Goldstone boson.  
These loops will relate, for instance, a heavy-quark operator 
of the form ${\C^k_{j_1j_2}}/{\Lambda^2}
(\bar{\mu}\Gamma_k\tau)(\bar{U}_{j_1}\Gamma_k U_{j_2})$~ 
to a light-quark operator of the form
${\C^\l_{i_1i_2}}/{\Lambda^2}
(\bar{\mu}\,\Gamma_\l\,\tau)(\bar{d}_{i_1}\,\Gamma_\l\, d_{i_2})$\,.\,   
Since we have very strong bounds
on operators with light $d_i$ quarks, 
this gives a method of deriving bounds on heavy 
$U_j$ quarks, and in many cases will provide the only bounds.

We consider the diagrams of Fig.\,3(a) and (b), and derive
the quark-bilinear,
\begin{equation}
\delta \C^{ij}_k \(\bar{q}_i\,\Gamma_k\,q_j\)  \,,
\end{equation}
where $\delta \C^{ij}_k$ is the loop-induced 
form factor computed from the diagrams.   
It is easy to show that
there will be no contribution to the scalar and 
pseudoscalar operators from the loop, and thus only
vector and axial-vector operators are relevant.

For $\Gamma_k=\gamma^\mu$, 
labelling the two external quarks with indices $(i_1,\,i_2)$ and the two
internal quarks with indices $(j_1,\,j_2)$, 
we find that the vector and axial-vector couplings
are generated with corresponding induced form factors,
\begin{equation}
\delta \C_V^{i_1,i_2} 
= -\delta \C_A^{i_1,i_2} 
= \frac{g^2|V_{i_1,j_1}V_{i_2,j_2}^*|}{32\pi^2}
\left(1-\frac{M_{j_1}^2M_{j_2}^2}{2M^2_W}\right)
\ln\frac{\Lambda}{M_W} \,,
\end{equation}
where $V_{ij}$ is the relevant CKM matrix element.
We note that if a mass-independent renormalization procedure
is used (cf. footnote in Ref.\,[59]), 
we would extract the leading logarithmic
term as $\ln\(\cut/\mu_0\)$ with $\mu_0$ set as the
energy scale of the relevant low energy process (e.g., 
$\mu_0 = m_b$ for $b$-decays).  But, for deriving the final bound on
$\cut$, this does not make much difference as it only goes
like $\sqrt{\ln (\cut/m_b)}$ [cf. Eq.\,(\ref{eq:cutWloop})
below]. Such difference would not be a main concern for leading logarithmic
estimates since in the leading logarithm approximation, all
unknown non-logarithmic terms are dropped by
assuming the absence of accidental cancellations.

For $\Gamma_k=\gamma^\mu\gamma^5$, 
we find the analogue of the above equation,
\begin{equation}
\delta C_V^{i_1,i_2} = -\delta C_A^{i_1,i_2} 
= -\frac{g^2|V_{i_1,j_1}V_{i_2,j_2}^*|}{32\pi^2}
\left(1+\frac{M_{j_1}^2M_{j_2}^2}{2M^2_W}\right)
\ln\frac{\Lambda}{M_W}\,.
\end{equation}
We will ignore small masses of light quarks and 
the overall signs are also irrelevant. 
So, the above leading logarithmic contribution 
is universal except when the internal loop
fields are both top quarks.   
It is useful that a single vector (or axial-vector) coupling 
induces both vector and axial-vector vertices. 
This allows us to use either 
the vector- or axial-vector type of light-quark bounds to
constrain both the vector- and axial-vector type of 
heavy-quark operators.   (Note also that in the limit where 
the muon mass is neglected, our $VV$ and $AA$ bounds from $\tau$ 
decays give identical bounds on $VA$ and $AV$.)

In particular, the scales  $\Lambda$ associated with heavy 
and light operators, respectively, can then be related. 
Letting $(X,\,Y)$ be the two internal quarks and $(x,y)$ be the two 
external quarks, we derive, in a transparent notation,
\begin{equation}
\label{eq:cutWloop}
\Lambda^{\pm}_{\mu\tau XY} 
= \dis\Lambda^{\pm}_{\mu\tau xy}
\sqrt{\frac{g^2}{32\pi^2}|V_{Xx}V_{Yy}^*|
\left(1\pm \frac{M_XM_Y}{2M^2_W}\right)
\ln\frac{\Lambda}{M_W}\, } ~,
\end{equation}
where the $+(-)$ sign is for the axial-vector (vector) coupling.  
Since, as discussed in Section 2, the loop cutoff $\cut$ in the 
leading logarithmic terms can reliably
represent the physical cutoff of the effective theory \cite{BL}, we may set
the above logarithmic cutoff $\cut$ equal to the light-quark bound 
$\cut^\pm_{\mu\tau xy}$.  
The light-quark bound $\cut^\pm_{\mu\tau xy}$
varies in the range around $1.5-15$\,TeV, 
and we may typically set it as 10\,TeV.

Now, we examine the vector- and axial-vector couplings for the 
$t-c$ and $t-u$ quark bilinears.   
With the internal quarks being the $t$ and $c$ (or $u$), 
the best bound comes from setting the external quarks to be $b$ (which
attaches to the internal $t$ line) and  $d$ 
(which attaches to the internal $c$ or $u$ line).  We then use the strong
axial-vector bound  $8.2$\,TeV for $b-d$ quark bilinear, 
to obtain the bounds
$310$\,GeV for the $t-c$ operator and $650$\,GeV for the $t-u$ operator
which hold for both axial-vector and vector type of couplings.
This bound is much stronger than that from the top decays.

\vskip 0.4cm
\noindent
\underline{\bf  {\Large $\bullet$} 
Charm Quark off-diagonal operators}
\vspace*{3mm}

The relevant charm quark operator is the $c-u$ operator.  One can obtain 
bounds from the loop corrections
discussed above for the vector and axial-vector 
operators.  There are several possible choices for the 
external lines (and
corresponding $ss$-vector, $dd$-vector or $sd$-axial-vector  bounds); 
the best comes from the vector $ss$-bound, i.e., from
$\tau\rightarrow \mu\phi$\,.  
We find that the bound on both the vector and axial-vector 
operators is \,$550$\,GeV.

The loop corrections will not give a significant bound on the scalar or 
pseudoscalar operators due to the chirality structure of the
couplings. We do get contributions from the
finite parts of the loop integrals, but 
the bounds are well below the W mass and thus not useful. 
For the pseudoscalar operator, one would get a strong bound from 
$D^0\rightarrow \mu\tau$, if it were kinematically
accessible, but it falls 20\,MeV short.
We have also considered virtual $\tau$ decays in
$D \rightarrow \mu \tau^\star 
\rightarrow \nu_\tau +$hadron, but this width is proportional 
to $\Gamma_\tau$ so that the realistically
obtainable experimental bound on  
$D \rightarrow \nu_\tau +$\,hadron will not give anything more 
than a few GeV bound for $\Lambda$. 
The scalar operator would 
require an additional pion in the final state,
which makes it even more inaccessible.  
We know of no bounds on the scalar and pseudoscalar operators.

\subsection{\large Diagonal Operators}

\vskip 0.3cm
\noindent
\underline
{\bf {\Large $\bullet$} 
$\mathbf{c}-\mathbf{c}$ Operators}
\vspace*{3mm}

The $c-c-\mu-\tau$ vector and axial vector operators can be bounded 
by the loop contributions (cf. Fig.\,3)
as discussed in the last subsection,
in which the internal quarks are both charm quarks.  The bound derived
from $\tau\rightarrow\mu\phi$ decays is
\,$\cut < 1.1$\,TeV,\, for both the vector and axial vector operators.
We note that there is no experimental bound available yet
for the decay \,$J/\Psi\to \mu\tau$\,.

One could bound the scalar and pseudoscalar operators by looking 
at $\mu\tau$ final state of $\chi_c$ and $\eta_c$ decays, 
respectively. However, no experimental bounds on these decays 
are available yet.

\vskip 0.4cm
\noindent
\underline{\bf {\Large $\bullet$} 
$\mathbf{b}-\mathbf{b}$ Operators}
\vspace*{3mm}

The obvious systems to look at are the $b\bar b$ bound states.
No experimental bound on $\Upsilon\rightarrow\mu\tau$ has been 
published. The ratio of the decay $\Upsilon(1s)\rightarrow\mu\tau
$ through the vector operator to the decay 
$\Upsilon(1s)\rightarrow\mu^+\mu^-$ is independent of the 
matrix element (if the mass difference between $\mu$ and $\tau$
is neglected) and is given by $144\pi^2M_\Upsilon^4/(e^4\Lambda^4)$.  
The $\Upsilon(1s)\rightarrow\mu^+\mu^-$ branching
ratio is $2.5\%$ \cite{upsilontautau}.
The upper bound on
$\Upsilon(1s)\rightarrow\mu\tau$ can be estimated by using 
Ref.\,\cite{upsilontautau} which
measured $\Upsilon\rightarrow\tau\tau$, and by comparing with the 
measurement of $\Upsilon(1s)\rightarrow\mu^+\mu^-$.  If one
assumes universality, these will be equal.  One can see that the 
excess of $\tau\tau$ events must be less
than about $0.40\%$  at $95\%$ C.L.
One then asks what fraction of $\mu\tau$ 
events would pass the cuts of the $\tau\tau$ analysis.  The $\tau\tau$
analysis selects events with one $\tau$ decaying to $e \nu \bar\nu$ 
and the other decaying to one-prong non-electron final states; this would 
be satisfied by $\mu\tau$ events depending on the cut on the momentum of 
the non-$e$ track.    
A conservative estimate\,\cite{urheimfirst} gives an upper bound of 
$4\%$ on the branching ratio, so that we arrive at 
\,$\Lambda > 180$\,GeV.\,   
We know of no bound on the scalar, pseudoscalar, or axial
vector couplings, since there is very little data on these bound states.

The loop contributions are negligible in this case,
primarily because the  CKM matrix elements for $b$
to $c,u$ transitions are small.

\vskip 0.4cm
\noindent\underline{\bf {\Large $\bullet$} 
$\mathbf{t}-\mathbf{t}$ Operators}
\vspace*{3mm}

The only possible way to bound the $t-t-\mu-\tau$ operators 
is through the loop discussed above, turning the top quarks
into $b,\ s$ or $d$.  The best
bound comes from the case in which the external quarks 
are $b$ and $d$, leading to $B^0\to\mu\tau$.  
Due to the small CKM matrix elements such as $V_{td}$, 
the bounds are not very strong.  For the axial 
vector operator, we find that $\Lambda > 115$\,GeV, and
for the vector operator, $\Lambda > 75$\,GeV.   
These bounds are below the mass of top quark and thus the 
effective theory description is no longer valid.   
As discussed below Eq.\,(III.55) for the constraints from 
top quark decays, the NDA analysis does increase 
the bounds by about a factor of $\sqrt{4\pi}\simeq 3.5$, but
even in this case the effective theory formalism may not be
so reliable.  Nevertheless, these weak bounds (if they might be
meaningful at all) are the best ones 
which we could obtain at the present.  
For the scalar and pseudoscalar operators, 
we know of no reasonable bounds at all.

\vskip 0.4cm
\noindent\underline{\bf {\Large $\bullet$} 
Radiative $\tau$ decays}
\vspace*{3mm}

One might expect that strong bounds on diagonal operators could be obtained
from $\tau\rightarrow\mu\gamma$, where the two quarks come together to form a loop.
If the photon is attached to the quark loop, the result for an on-shell
photon vanishes. But, if the photon is attached to the tau or muon
line, the quark loop is quadratically divergent and independent
of external momentum.  As discussed in Sec.\,IID, 
however, quadratically divergent corrections
are not guaranteed to represent the real contributions of a heavy physical
state, and may be absorbed via renormalization, and thus will not be
considered further.

\vspace*{5mm}
\section{Summary}
\vspace*{-1mm}

The strong experimental evidence for
large $\nu_\mu - \nu_\tau$ oscillation motivates us
to explore the allowed mixings 
between the second and third generations in the charged
lepton sector. 
We have systematically analyzed bounds 
on the generic dimension-6 flavor-violation
$\tau-\mu$ operators of the form 
$$
(\bar\mu \,\Gamma\, \tau)( \bar{q}^\alpha \,\Gamma\, {q}^\beta)\,,
\label{fourfermiop}
$$
where the Dirac matrices $\Gamma \in (S,P,V,A)$ 
are the same for both the 
$\tau-\mu$ and the $ q^\alpha - q^\beta$ bilinears.
Such effective operators are interesting as they can naturally arise  
from various new physics scenarios and reflect the underlying 
flavor-mixing dynamics which may directly link to the 
large  $\nu_\mu - \nu_\tau$ neutrino oscillation.
Since the neutrino oscillations measure the MNS mixing matrix
which is only a product of two  rotation matrices from the
lepton and neutrino mass-diagonalizations 
[cf. Eq.\,(\ref{eq:MNS})],
it is thus important to fully understand the flavor dynamics
and the origin of the neutrino oscillation phenomena
by directly testing possible large lepton mixings 
from additional flavor-violation processes.
Given such generic dimension-six $\tau-\mu$ effective operators, 
we have considered 
all possible flavor combinations in the quark-bilinear 
$\bar{q}^\alpha \,\Gamma\, {q}^\beta$ and analyzed 
existing experimental data for a variety of processes to establish 
the best available bounds on the invoked new physics scale
$\Lambda$.  Our results are summarized in Table\,\ref{table1}.

\begin{table}
{\large 
\begin{center}
\begin{tabular}{c|cccc}
\hline \hline
& & & & \\[-2mm]
  ~~Bound~~ & $ 1 $ & $ \gamma^{~}_5$ & $\gamma_\sigma$ & 
              $\gamma_\sigma \gamma^{~}_5 $ \\
& & & & \\[-2mm]
\hline\hline
& & & & \\[-3mm]
$\bar{u} u$ & 2.6 TeV & 12 TeV & 12 TeV & 11 TeV \\           
& ~($\tau \rightarrow \mu \pi^+ \pi^-$)~ & 
~$(\tau \rightarrow \mu \pi^0$)~ & 
~($ \tau \rightarrow \mu \rho $)~ & 
~($ \tau \rightarrow \mu \pi^0 $)~  \\
& & & & \\[-3mm]
\hline
& & & & \\[-3mm]
$\bar d d$ & 2.6 TeV & 12 TeV & 12 TeV & 11 TeV \\
           & ($ \tau \rightarrow \mu \pi^+ \pi^-$) & ($\tau \rightarrow \mu \pi^0$) & ($ \tau \rightarrow \mu \rho $) & ($ \tau \rightarrow \mu \pi^0 $) \\ 
& & & & \\[-3mm]
\hline
& & & & \\[-3mm]
$\bar s s$ & 1.5 TeV & 9.9 TeV & 14 TeV & 9.5 TeV \\
           & ($ \tau \rightarrow \mu K^+ K-$) & ($\tau \rightarrow \mu \eta$) & ($ \tau \rightarrow \mu \phi $) & ($ \tau \rightarrow \mu \eta $) \\ 
& & & & \\[-3mm]
\hline
& & & & \\[-3mm]
$\bar s d$ & 2.3 TeV & 3.7 TeV  & 13 TeV & 3.6 TeV  \\
           & ($ \tau \rightarrow \mu K^+ \pi^-$) & ($\tau \rightarrow \mu K^0$) & ($ \tau \rightarrow \mu K^\star $) & ($ \tau \rightarrow \mu K^0 $) \\ 
& & & & \\[-3mm]
\hline
& & & & \\[-3mm]
$\bar b d$ & 2.2 TeV  & 9.3 TeV  & 2.2 TeV  & 8.2 TeV  \\
           & ($ B \rightarrow \pi \mu \tau $) & ($B \rightarrow \mu \tau$) & ($ B \rightarrow \pi \mu \tau $) & ($ B \rightarrow \mu \tau $) \\ 
& & & & \\[-3mm]
\hline
& & & & \\[-3mm]
$\bar b s$ & 2.6 TeV  & 2.8 TeV  & 2.6 TeV  & 2.5 TeV  \\
           & ($ B \rightarrow K \mu \tau $) & ($B_{\rm s} \rightarrow \mu \tau $) & ($ B \rightarrow K \mu \tau $) & ($ B_{\rm s} \rightarrow \mu \tau $) \\ 
& & & & \\[-3mm]
\hline
& & & & \\[-3mm]
$\bar t c$ & 190 GeV  & 190 GeV  & 310 GeV & 310 GeV  \\
           & ($ t \rightarrow c \mu \tau $) & ($t \rightarrow c \mu \tau $) 
& ($B \rightarrow  \mu \tau$) & ($B \rightarrow  \mu \tau $) \\ 
& & & & \\[-3mm]
\hline
& & & & \\[-3mm]
$\bar t u$ & 190 GeV  & 190 GeV  & 650 GeV & 650 GeV  \\
           & ($t \rightarrow u \mu \tau $) & ($t \rightarrow u \mu \tau $) & ($ B \rightarrow \mu \tau $) & ($ B \rightarrow \mu \tau $) \\ 
& & & & \\[-3mm]
\hline
& & & & \\[-3mm]
$\bar c u$ & $\star$ & $\star$  & 550 GeV & 550 GeV  \\
           &  &  & ($ \tau \rightarrow \mu \phi $) & ($ \tau \rightarrow \mu \phi $) 
\\ 
& & & & \\[-3mm]
\hline
& & & & \\[-3mm]
$\bar c c$ & $\star$ & $\star$  & 1.1 TeV & 1.1 TeV  \\
           &  &  & ($ \tau \rightarrow \mu \phi $) & ($ \tau  \rightarrow \mu \phi $)
\\ 
& & & & \\[-3mm]
\hline
& & & & \\[-3mm]
$\bar b b$ & $\star$ & $\star$ & $ 180$ GeV & $\star$  \\
           &  &  &  ($\Upsilon\rightarrow \mu \tau$)  &  \\ 
& & & & \\[-3mm]
\hline
& & & & \\[-3mm]
$\bar t t$ & $\star$ & $\star$  & 75 GeV & 120 GeV  \\
           &  &  & ($ B \rightarrow \mu \tau $) & ($ B \rightarrow \mu \tau $) 
\\[1.5mm]
\hline\hline
\end{tabular}
\end{center}
}
\caption{Bounds at 90\%\,C.L. on four-Fermi flavor-violation operators 
of the form $(\bar\mu\, \Gamma_j\tau )(\bar{q}^\a\, \Gamma_j \,q^\b)$,
where $\Gamma_j \in ({\rm S, P, V, A})$. Combinations for which no 
bound has been found are marked with an asterisk,
otherwise we list the process which gives the strongest bound 
(cf. text for details).}
\label{table1}
\end{table}

In Section IIIA, we studied the operators in 
Eq.\,(\ref{eq:taumu6}) involving two light quarks.   
We found quite strong bounds in this case since 
there are good experimental limits on decays of 
$\tau$ to $\mu$ and one or two light mesons.  
For $P$, $V$ and $A$ the bounds on $\Lambda$ are of order 
$ 10$\,TeV (except for $\bar s d$ 
where the $P$ and $A$ bounds are weaker since the
experimental constraint is weaker).  
Since three-body decays are used 
for the $S$ case, the bounds obtained are slightly smaller, 
of order $1.5 - 2.5$\,TeV.

In Sections IIIB and IIIC, we studied operators 
involving at least one heavy quark.   
For the case of a charm quark in 
Eq.\,(\ref{eq:taumu6}), one might think of 
using $D$-decays to final states involving $\mu \tau$
 but these are ruled out by kinematics.  
However, it turns out that we obtained good bounds on 
operators involving the $cu$ and $cc$ combinations 
for the $V$ and $A$ cases by considering
loop contributions (shown in Fig.\,3) to $\tau \rightarrow \mu \phi$.  
Loop contributions involving $c$ quarks are 
enhanced by the fact that $V_{cs} \approx 1$.  
For the $S$ and $P$ cases we could not find a bound on 
$\Lambda$ since the loop diagrams shown in Fig.\,3 
have vanishing leading logarithmic term and the remaining
finite loop terms are numerically negligible.  
Also there is no bound from pseudoscalar 
and scalar charmonium decays --- these states are 
rather broad with several MeV uncertainties in their 
widths at the present, so there is no significant experimental 
constraint on their branching fractions to $\mu \tau$.

For operators involving one $b$-quark, 
we used the experimental limits on $B \rightarrow \mu \tau$, 
$B_s \rightarrow \mu \tau$ and estimated 
bounds on $B \rightarrow \mu \tau M$ \footnote{$M$ denotes a light
pseudoscalar meson.}\, 
to obtain bounds on $\Lambda$ in the $3-9$ TeV range.
There is some uncertainty in our approximate 
HQET\footnote{HQET stands for Heavy Quark Effective Theory.} 
hadronic matrix elements which could be 
improved, but since it is a fourth root which
appears in our extraction of $\Lambda$, the final error is 
not so large.  For the $\bar b b $ case, we considered 
the contribution of loop diagrams in Fig.\,3 
with internal $b$ quarks to processes involving external $u$ 
quarks.  However, these are suppressed by a CKM factor 
$ {\left| V_{ub} \right| }^2$ and thus do not lead to a 
useful bound.  The only significant bound  for $\bar b b $ which
we could obtain is for the $V$-type operator 
and it comes from $\Upsilon$ decay.

For operators involving one top quark (via $tu$ and $tc$), 
the best bounds for the $S$ and $P$ cases come directly from 
the top quark decays 
$ t \rightarrow c \mu \tau, u \mu \tau$.  
Their decay widths scale as
\,${m_t}^5/\Lambda^4$\, so that the bounds are actually non-trivial.  
For the $V$ and $A$ types of 
operators with quark bilinears $tu$, $tc$ and also $\bar t t$, 
the strongest bound comes from 
the associated contribution to $ B\rightarrow \mu \tau$ 
due to internal $t$ quarks in Fig.\,3.  
We have not found a way to bound the $S$ and $P$ 
operators for the case with $\bar t t$ quark bilinear.

It is important to note that under the linear realization of 
the Standard Model gauge group, the allowed operators are 
restricted to $V$ and $A$ types, as discussed in 
Sec.\,IIA and Appendix\,A.  
In this case, we obtain bounds for all but one of the allowed 
operators  (i.e., except $\GA = A$ for the quark-bilinear 
$\bar{b}\,\GA\, b$).  
We should mention that if the coefficients $\C^j_{\a\b}$
of our dimension-6
operators follow the estimate of the NDA analysis 
in Eq.\,(\ref{eq:C-NDA}) for certain class of strongly coupled
theories [instead of the ``default'' estimate in  Eq.\,(\ref{eq:C-NDA})], 
the final bounds in Table\,I would be stronger by a factor of
$\lesssim \!\sqrt{4\pi} \simeq 3.5$.  
On the other hand, if the underlying theory
is weakly coupled (such as SUSY-type models), the  
coefficients $\C^j_{\a\b} \sim \O(1)$ [cf. Eq.\,(\ref{eq:C-Weak})]
and thus the final bounds in
Table\,I would be weaker by about a factor of
$\sqrt{4\pi}$\,. 
Impressively, even in this weakly coupled scenario,
Table\,I shows that significant
bounds of \,$\cut > 0.6-4$\,TeV\,
still hold for all those quark bilinears with no $t$ or $c$ quark
and at most one $b$ quark. Also, it will be interesting to further 
investigate how the present bounds in Table\,I can explicitly constrain
the relevant models classified in Sec.\,IIC,  which will induce
specific forms of our
$\mu-\tau$ flavor violation operators in the low energy theory.

We note that most bounds listed in Table\,I are from rare
decays of $\tau$ and $B$.  
Tighter experimental bounds on $\tau$ decays would lead 
to stronger bounds on almost one half of the
operators considered here. Searches for $\mu\tau$ from charmonium
decays ($J/\Psi,\chi_c,\eta_c$) would be an important addition to
study this class of operators.
Also, as emphasized in Section\,III, 
it would be particularly significant to 
have an experimental bound on $B_s \rightarrow \mu \tau$ 
and also $B$ decays to $\mu \tau M$.
Since we have not found any way to bound three of the four 
operators with
$\bar b b $ bilinears, it would be helpful to have 
experimental bounds on scalar and 
pseudoscalar $\bar{b}b$ decay to $\mu \tau$. 
It is also very interesting to search for the decay 
$t \rightarrow \mu\tau + {\rm jet}$ at the top-quark factories
such as the Tevatron Run-II and the CERN LHC.

Finally, in Appendix\,B, we also considered an extension of our
formalism in Sec.\,IIA to include purely leptonic $\tau-\mu$
operators at dimension-6 with a lepton-bilinear 
$(\ov{\l}^\a\,\GA\,\l^\b)$ [cf. Eq.\,(\ref{eq:taumu6l})].
Among the available constraints, we found that the three-body
rare decays $\tau\to 3\mu,\,\mu\mu e,\,\mu ee$ give the best
bounds at the order of $10$\,TeV or so.
For $(\ov{\l}^\a\,\GA\,\l^\b)$ being a neutrino pair, the bounds
from $\tau\to \mu\nu\bar{\nu}$ decay are much weaker, around
$2-3$\,TeV. No significant bound is obtained for 
$(\ov{\l}^\a\,\GA\,\l^\b)$ containing one or two heavy $\tau$
leptons.

\vspace*{15mm}
\section*{\large \hspace*{-6mm}Acknowledgments}
\vspace{-3mm}

We are very grateful to Jon Urheim for numerous discussions 
about $B$-decays and $\Upsilon$-decays, 
and to Jose Goity for useful discussions.   
D.B. acknowledges the support from the Thomas Jefferson
National Accelerator Facility operated by the Southeastern Universities
Research Association (SURA) under DOE contract No.~DE-AC05-84ER40150;
T.H. was supported in part by a US DOE grant 
No.~DE-FG02-95ER40896 and by the Wisconsin Alumni Research Foundation;
H.J.H. was supported by U.S. Department of Energy under
grant No.~DE-FG03-93ER40757;
and M.S. was supported by the National Science Foundation
through grant PHY-9900657.

\newpage
\appendix


\noindent
\section{Nonlinear versus Linear Realization of the
         Electroweak Gauge Symmetry}

We note that in principle the exact form of Eq.\,(\ref{eq:taumu6})
depends on how the electroweak gauge symmetry is realized in
$\LL_{\rm SM}$. The general form  (\ref{eq:taumu6}) can be derived
by using the nonlinear realization of the SM gauge symmetry.
Under the nonlinear realization, 
the SM Higgs-Goldstone fields are parameterized as
\beq
\label{eq:U}
\Phi = {1\over \sqrt{2}}(v+H)U \,,~~~~U = \exp [i\pi^a\tau^a/v]\,,
\eeq
which transforms, under \,$\GSM=SU(2)_L\otimes U(1)_Y$,\, as
\beq
\label{eq:nonlinear-transf}
\begin{array}{ll}
U\rightarrow U' = g_L^{~} U g_Y^{\dagger}\,, ~~&~~
H\rightarrow H' = H \,,\\[0.3cm]
g_L^{~} = \exp [-i\theta_L^a\tau^a/2]\,,  ~~&~~
g_Y^{~} = \exp [-i\theta_Y\tau^3/2] \,.
\end{array}
\eeq
We introduce the following useful notations,
\beq
\begin{array}{l}
\dis
D_\mu U =\partial_\mu U +igW^a_\mu{\tau^a\over 2}U-ig'UB_\mu 
{\tau^3\over 2} \,,
\\[3mm]
{\cal V}_\mu = (D_\mu U)U^{\dagger}    \,,~~~~
\ov{\cal V}_\mu = U^{\dagger}(D_\mu U)
                =U^\dag{\cal V}_\mu U \,,~~~~
{\cal T}^a = U\tau^a U^{\dagger}       \,,  
\\[3mm]
\dis
\W_\mu^\pm 
        = -{i\over g}{\rm Tr}[\tau^\pm U^{\dagger}D_\mu U]
        = -{i\over g}{\rm Tr}[\tau^\pm \overline{\cal V}_\mu ]  
        = -{i\over g}{\rm Tr}[{\cal T}^\pm {\cal V}_{\mu}] \,,
\\[3mm]
\dis
\Z_\mu^0
        = -{i\cw\over g}{\rm Tr}[\tau^3 U^{\dagger}D_\mu U]
        = -{i\cw\over g}{\rm Tr}[\tau^3 \overline{\cal V}_\mu ]  
        = -{i\cw\over g}{\rm Tr}[{\cal T}^3 {\cal V}_{\mu}] \,,
\end{array}
\eeq
where $~\cw\equiv \cos\theta_W~$ and 
      $~\sw\equiv \sin\theta_W\,$.\,
It can be proven that
$\,
\ov{\cal V}_\mu \,=\, i\dis\f{g}{2}
\[\W^+_\mu \tau^- + \W^-_\mu \tau^+  + \cw^{-1}\Z_\mu^0\tau^3\]
$.\,
Then, we can write the following transformation laws, under $\GSM$,
\beq
\label{eq:transf}
\begin{array}{l}
{\cal V}_{\mu}\rightarrow {\cal V}_{\mu}^{\prime}  
= g_L^{~} {\cal V}_{\mu} g_L^{\dagger}\,,~~~~
\ov{\cal V}_{\mu}\rightarrow \overline{\cal V}_{\mu}^{\prime}  
= g_Y^{~} \overline{\cal V}_{\mu} g_Y^{\dagger}\,,~~~~
\\[3mm]
\W_{\mu}^{\pm}\rightarrow
\W_{\mu}^{\pm \prime} =  \exp [\mp i\theta_Y^{~}] \W_{\mu}^{\pm} \,,~~~~
\Z_{\mu}^{0}\rightarrow
\Z_{\mu}^{0 \prime}  = \Z_{\mu}^{0} \,.
\end{array}
\eeq
Since $~(\W^\pm,\,\Z^0)$ 
feel only the unbroken $U(1)_{\rm em}$ gauge interaction, 
its covariant derivative is
\beq
D_\nu \W^\pm_\mu = (\partial_\nu \pm ieA_\nu )\W^\pm_\mu  \,,~~~
D_\nu \Z^0_\mu   = \partial_\nu \Z^0_\mu  \,,
\eeq
In general, the non-linear composite fields 
$~(\W^\pm,\,\Z^0)~$ can be expanded as
\beq
\begin{array}{l}
\displaystyle
\W^\pm_\mu = W^\pm_\mu +
\frac{1}{M_W}\partial_\mu\pi^\pm\pm
\frac{ig}{M_W}\left[
\({\rm c}_{\rm w}Z_\mu + {\rm s}_{\rm w}A_\mu\)\pi^\pm
-W^\pm_\mu\pi^0\right]
\pm\f{ig}{2M_W^2}\[\pi^0\partial_\mu\pi^\pm -
                 \pi^\pm\partial_\mu\pi^0 \]
+\cdots    ,
\\[4mm]
\displaystyle
\Z^0_\mu = Z_\mu^0 +
\frac{1}{M_Z}\partial_\mu \pi^0 +\frac{ig}{M_Z}
\(W^+_\mu\pi^- -W^-_\mu\pi^+ \) +
\f{ig}{2{\rm c}_{\rm w} M_Z^2}
\(\pi^+\partial_\mu\pi^--\pi^-\partial_\mu\pi^+ \)
+\cdots  ,
\end{array}
\eeq
so that in the unitary gauge, 
\,$(\W_\mu^\pm,\,\Z^0_\mu)=(W_\mu^\pm,\,Z^0_\mu)$,\,
and 
\beq
{\cal V}_\mu =\ov{\cal V}_\mu =\dis\frac{ig}{2}
\left\lgroup
\begin{array}{ll} 
c^{-1}_{\rm w}Z_\mu & \sqrt{2}W^+_\mu 
\\[3mm] 
\sqrt{2}W^- & -c^{-1}_{\rm w}Z_\mu \end{array} \right\rgroup .
\eeq
Now, we can rewrite the SM Lagrangian in terms of fields 
which feel only the unbroken $U(1)_{\rm em}$, 
\beq
\label{eq:LnSM}
\begin{array}{rl}
{\cal L}_{\rm SM} & = {\cal L}^G_{\rm SM} +{\cal L}^H_{\rm SM} 
                 + {\cal L}^F_{\rm SM}
\\[3mm]
\displaystyle 
{\cal L}^G_{\rm SM} & =  
             -{1\over 4}\[{\W}^a_{\mu\nu}{\W}^{a\mu\nu}
                         +B_{\mu\nu}B^{\mu\nu}   \]
\\[3mm]
\displaystyle
{\cal L}^H_{\rm SM} 
& = {1\over 2}\partial_\mu H\partial^\mu H - V_{\rm SM}(H) 
            -{1\over 4}{\rm Tr}
             \left[{\cal V}^\mu {\cal V}_\mu\right]
             \left[ v^2+2vH+H^2 \right]
\\[3mm]
\displaystyle
{\cal L}^F_{\rm SM} 
 & = \displaystyle
-\f{g}{2}
\overline{F_{jL}} \gamma^\mu
 \(\W_\mu^\pm\tau^\mp+{\cw^{-1}}\Z^0_\mu\tau^3\) F_{jL}
+\overline{f_{j}} i\gamma^\mu 
\[\partial_\mu + ieQ_{f_j} A^0_\mu 
               - ig'\sw Q_{f_j} \Z^0_\mu \]f_j 
\\[3mm]
& \hspace*{4mm}
-\overline{f_{j}}M^f_{jj'}f_{j'}\left(1+{H}/{v}\right)
\end{array}
\eeq
where $~M^f~$ is the general fermion mass-matrix and after 
diagonalization, $~M^f_{jj'}=m_{fj}\delta_{jj'}\,$.\,
The electric charge of the fermion $~f_j~$ is defined by
$~Q_{f_j}=I_{3j}+ {Y_{f_j}}/{2} \,$,\, 
and $~F_{jL}\equiv (f_{1jL}, f_{2jL})^T~$ is the left-handed
$SU(2)_L$ fermion doublet.  
For the electroweak interactions in this non-linear realization,
Eqs.\,(\ref{eq:LnSM}) and (\ref{eq:transf})
show that the fermions have the $U(1)_{\rm em}$ covariant derivative
\beq
D_\mu f_j = \left(\partial_\mu +ieQ_{f_j}A_\mu\right)f_j \,.
\eeq
Since the fermion fields only feel an unbroken electromagnetic
$U(1)_{\rm em}$ gauge symmetry,
we see that under this formalism the dimension-six
$\tau-\mu$ operator contained in $\Delta{\cal L}$ indeed 
takes the most general form as in Eq.\,(\ref{eq:taumu6}) 
that involves one $\tau-\mu$ bilinear and one quark bilinear.
This nonlinear formalism is particularly motivated when the Higgs sector of
$\LL_{\rm SM}$ is strongly coupled or the Higgs boson 
does not exist \cite{Hill-Rept,App}.
In this case, the electroweak symmetry breaking  
scale is bounded from the above, i.e.,
$\cut_{\rm EW} \lesssim 4\pi v$ \cite{weinberg,georgi1,georgi2}.

If the Higgs boson $H^0$ is relatively light, we may choose
the linear realization for the $\LL_{\rm SM}$ \cite{BW}, 
in which we consider
$\cut \geq \cut_{\rm EW}$ so that at the new physics scale $\cut$, the
effective Lagrangian $\LL_{\rm eff}$ should be invariant under SM gauge group 
$\GSM=SU(2)_L\otimes U(1)_Y$. 
The SM fermion fields in these two formalisms are related \cite{PZ},
\beq
\begin{array}{l}
F_{jL} = U^\dagger F_{jL}^{\rm linear}\,,~~~~ 
f_{jR} = f_{jR}^{\rm linear} \,,
\\[3mm]
f_{j}\rightarrow f_{j}^{\prime} = \exp [i\theta_Y Q_{f_j}]f_j \,,
~~~~({\rm under~}\GSM) \,,
\\[2mm]
\end{array}
\eeq
where the linearly realized fermions have the usual covariant
derivatives as in the SM,
\beq
\begin{array}{l}
\displaystyle
D_\mu F_{jL}^{\rm linear} 
= \left(\partial_\mu +ig{\tau^a\over2}W^a_\mu+
  ig'{Y_{jL}\over2}B_\mu\right)F_{jL}^{\rm linear}\,,~~~~
\dis
D_\mu f_{jR}^{\rm linear}
=  \left(\partial_\mu +ig'{Y_{jR}\over2}B_\mu\right)f_{jR}^{\rm linear}\,.
\end{array}
\eeq
This further restricts the form of the dimension-6 operator  
in Eq.\,(\ref{eq:taumu6}) because of the requirement of both the
isospin and hypercharge conservations. Thus, we arrive at
Eqs.\,(\ref{eq:taumu6-lin}) and (\ref{eq:taumu8-lin})
for the linearly realized effective $\tau-\mu$
operators in Sec.\,IIA.

\noindent
\section{Bounds on the Purely Leptonic Operators}

In principle, we can extend the formalism in Sec.\,II to
include the purely leptonic operators of the form,
\beq
\label{eq:taumu6l}
\DL^{(6\l )}_{\tau\mu} 
=\dis\sum_{j,\alpha,\beta}
\f{~\C^\l_{j,\alpha\beta}~}{\cut^2}
\left( \ov{\mu} ~\Gamma_j\, \tau \right) 
\left( \ov{\l}^\alpha \,\Gamma_j\, \l^\beta \right)
\,+\,{\rm H.c.}\,,
\eeq
which contains an additional lepton bilinear 
$(\ov{\l}^a\, \Gamma\, \l^\b )$
instead of quark bilinear $(\ov{q}^a\, \Gamma\, q^\b )$
in Eq.\,(\ref{eq:taumu6}).

For both $\l^a$ and $\l^b$
being light leptons (electrons or muons),
the three-body rare $\tau$ decay,
$\tau^- \to \mu^- l^\alpha \ov{l}^\beta$, provides the best bounds.  
There are in total four allowed combinations from 
Eq.\,(\ref{eq:taumu6l}), 
$(\l^\a,\,\ov{\l}^\b) = 
(\mu^-,\,\mu^+),\,
(\mu^-,\,e^+),\,         
(\mu^+,\,e^-),\,
(e^-,\,e^+)
$, giving the decay channels
$\tau^-\to \mu^-\mu^-\mu^+,\, \mu^-\mu^-e^+,\,
\mu^-\mu^+e^-,\, \mu^-e^-e^+$.
Their partial decay widths are computed as
\beq
\GA_{3\l} =\dis
\f{~(\C_{\a\b}^{\l,j})^2\,\varrho\,\, m_\tau^5~}{768\pi^3\cut^4}\times
\left\{
\ba{ll} \dis 1,         &  ~~~~(\GA_j =S,\,P), \\[2mm]
             4,         &  ~~~~(\GA_j =V,\,A),
\ea\right.
\eeq
where \,$\varrho = 1$\, for 
\,$\tau^-\to \mu^-\mu^-\mu^+,\, \mu^-\mu^-e^+$,\,
and \,$\varrho = 1/2$\, for
\,$\tau^-\to \mu^-\mu^+e^-,\, \mu^-e^-e^+$.

The experimental bounds on the decay branching ratios
of these channels are given, at  90\%\,C.L. \cite{pdg},
\beq
\label{eq:BR-3l}
\ba{llllll}
{\rm Br}[\tau^-\to \mu^-\mu^-\mu^+] \!\!&<&\!\! 1.9\times 10^{-6}, &
~~~{\rm Br}[\tau^-\to \mu^-\mu^-  e^+] \!\!&<&\!\! 1.5\times 10^{-6},
\\[2mm]
{\rm Br}[\tau^-\to \mu^-\mu^+ e^-] \!\! &<&\!\! 1.8\times 10^{-6}, &
~~~{\rm Br}[\tau^-\to \mu^- e^-  e^+]  \!\!&<&\!\! 1.7\times 10^{-6},
\ea
\eeq
which are all at $10^{-6}$ level.
From these we derive the following bounds on the scale $\cut$,
\beq
\label{eq:bound-ll}
\ba{lcl}
\cut & > &\dis \left\{
\ba{ll}
(12.8,\,13.5,\,11.9,\,11.0)\,{\rm TeV},  & ~~~(\GA_j=S,\,P),
\\[2mm]
(18.0,\,19.1,\,15.4,\,15.6)\,{\rm TeV},  & ~~~(\GA_j=V,\,A),
\ea
\right.
\ea
\eeq
for the four rare  decay channels,
$\tau^-\to \mu^-\mu^-\mu^+,\, \mu^-\mu^-e^+,\,
\mu^-\mu^+e^-,\, \mu^-e^-e^+$, respectively.
We see that these are quite similar, around $11-14$\,TeV
for $\GA_j=S,\,P$ and $15-19$\,TeV for  $\GA_j=V,\,A$.

For the case where $\l^\alpha$ and $\l^\beta$ are neutrinos, 
$(\l^\a,\,\l^\b)=(\nu^\a,\,\nu^\b)$,
then the decay rate for 
$\tau\rightarrow\mu \nu\bar\nu$ will be increased.  
Given the current data \cite{pdg},
\,${\rm Br}[\tau\to \mu\nu\bar\nu] =(17.37\pm 0.07)\%$,\,
we see that an increase of this branching ratio by an amount of
$0.115\%=1.15\times 10^{-3}$ can be tolerated at 90\%\,C.L. 
Similar to Eq.\,(\ref{eq:bound-ll}), we derive the following
bound from the $\tau\rightarrow\mu \nu^\a \ov{\nu}^\b$ channel,
\beq
\label{eq:bound-nn}
\ba{lcl}
\cut & > &\dis \left\{
\ba{ll}
2.2\,{\rm TeV},  ~~~& ~~~(\GA_j=S,\,P),
\\[2mm]
3.1\,{\rm TeV},  ~~~& ~~~(\GA_j=V,\,A),
\ea
\right.
\ea
\eeq
which are weaker than (\ref{eq:bound-ll}) by about a factor of
~$(1000)^{1/4}\approx 5- 6$~ due to the different branching ratio.

Finally, one or both of $(\l^a,\,\l^\b)$
can be the $\tau$ lepton.  In this case, we may use the triangle
$W$-loop to relate the heavy $\tau$ lepton operators 
to the neutrino operators, similar to Fig.\,3 and 
Eq.\,(\ref{eq:cutWloop}).
The resulting bounds are around
$150$\,GeV for the $S$ and $P$ operators, and around
$200$\,GeV for the $V$ and $A$ operators, 
which are rather weak and less useful.

\newpage
\renewcommand{\baselinestretch}{1.4}

\end{document}